\documentclass[amsmath,amssymb,aps,longbibliography, superscriptaddress]{revtex4-2}

\usepackage{bbm}
\usepackage{multirow}
\usepackage{float}
\usepackage{graphicx} 
\usepackage{xcolor}

\usepackage{changes}
\definecolor{avicolor}{RGB}{216, 27, 96}

\definecolor{riverlane_green}{RGB}{0, 150, 143}
\definecolor{riverlane_orange}{RGB}{255, 117, 0}

\usepackage{hyperref}
\hypersetup{
  colorlinks   = true, 
  urlcolor     = riverlane_green, 
  linkcolor    = riverlane_green, 
  citecolor   = riverlane_orange  
}

\definecolor{avicolor}{RGB}{216, 193, 7}
\definechangesauthor[name={Aleksei}, color={avicolor}]{AVI}

\usepackage{braket}
\usepackage[compat=0.6]{yquant}

\newcommand{\tr}{ {\bf r}}
\newcommand{\tx}{ {\bf x}}

\newcommand{\tk}{ {\bf k}}
\newcommand{\tq}{ {\bf q}}
\newcommand{\ta}{ {\bf a}}
\newcommand{\tb}{ {\bf b}}
\newcommand{\tR}{ {\bf R}}
\newcommand{\tT}{ {\bf T}}

\newcommand{\tG}{ {\bf G}}
\newcommand{\tn}{ {\bf n}}

\newcommand{\hc}{ {\hat c}}
\newcommand{\ha}{ {\hat a}}

\newcommand{\IINT}[1]{\iint\limits_{#1} d\tx \, d\tx'\,}
\newcommand{\INTr}[1]{\int\limits_{#1} d^3 \tr\,}
\newcommand{\IINTr}[1]{\iint\limits_{#1} d^3\tr \, d^3\tr'\,}

\newcommand{\adps}{\hat a^{\dagger}_{p\sigma}}

\newcommand{\aps}{\hat a_{p\sigma}}
\newcommand{\aqs}{\hat a_{q\sigma}}

\newcommand{\adrt}{\hat a^{\dagger}_{r\tau}}

\AtBeginDocument{\renewcommand{\ast}{\hat a_{s\tau}}} 

\newcommand{\pso}{\hat \gamma_{p\sigma, 0}}

\newcommand{\psn}{\hat \gamma_{p\sigma, 1}}

\newcommand{\psj}{\hat \gamma_{p\sigma, j}}
\newcommand{\qsj}{\hat \gamma_{q\sigma, j}}
\newcommand{\rsj}{\hat \gamma_{r\sigma, j}}
\newcommand{\ssj}{\hat \gamma_{s\sigma, j}}

\newcommand{\qtj}{\hat \gamma_{q\tau, j}}

\newcommand{\psk}{\hat \gamma_{p\sigma, k}}
\newcommand{\qsk}{\hat \gamma_{q\sigma, k}}
\newcommand{\rsk}{\hat \gamma_{r\sigma, k}}
\newcommand{\ssk}{\hat \gamma_{s\sigma, k}}

\newcommand{\rtk}{\hat \gamma_{r\tau, k}}
\newcommand{\stk}{\hat \gamma_{s\tau, k}}
\newcommand{\rtl}{\hat \gamma_{r\tau, l}}
\newcommand{\stl}{\hat \gamma_{s\tau, l}}
\newcommand{\stm}{\hat \gamma_{s\tau, m}}

\begin{document}

\title{Quantum Computation for Periodic Solids in Second Quantization}

\author{Aleksei V. Ivanov}
\email[Corresponding author: ]{aleksei.ivanov@riverlane.com}
\author{Christoph S\"underhauf}
\author{Nicole Holzmann}
\affiliation{Riverlane, St Andrews House, 59 St Andrews Street, Cambridge, CB2 3BZ, United Kingdom}

\author{Tom Ellaby}
\author{Rachel N. Kerber}
\author{Glenn Jones}
\affiliation{Johnson Matthey Technology Centre, Blounts Court, Sonning Common, RG4 9NH, United Kingdom}
\author{Joan Camps}
\affiliation{Riverlane, St Andrews House, 59 St Andrews Street, Cambridge, CB2 3BZ, United Kingdom}

\begin{abstract}
In this work, we present a quantum algorithm for ground-state energy calculations of periodic solids on error-corrected quantum computers. The algorithm is based on the sparse qubitization approach in second quantization and developed for Bloch and Wannier basis sets.
We show that Wannier functions require less computational resources with respect to Bloch functions because: (i) the L$_1$ norm of the Hamiltonian is considerably lower and (ii) the translational symmetry of Wannier functions can be exploited in order to reduce the amount of classical data that must be loaded into the quantum computer. 
The resource requirements of the quantum algorithm are
estimated for periodic solids such as NiO and PdO. These transition metal oxides are industrially relevant for their catalytic properties. We find that ground-state energy estimation of Hamiltonians approximated using 200--900 spin orbitals requires {\it ca.}~$10{}^{10}$--$10^{12}$ T gates and up to $3\cdot10^8$ physical qubits for a physical error rate of $0.1\%$.
\end{abstract}

\maketitle

\section{Introduction}

Quantum mechanical simulation of molecules and materials is a promising application area of quantum computers~\cite{liu_prospects_2022, bauer_quantum_2020, mcardle_quantum_2020} that will enable the calculation of key properties of chemical systems with controllable errors using physically accurate models. 
Following Feynman's original idea of modelling quantum systems on quantum computers~\cite{feynman_simulating_1982} and the first formalized procedures for carrying out such simulations~\cite{abrams_simulation_1997,abrams_quantum_1999,ortiz_quantum_2001}, a plethora of quantum algorithms for calculating energies of molecular systems have been developed in recent years
~\cite{aspuru_guzik_simulated_2005, kassal_polynomial_time_2008, whitfield_simulation_2011, seeley_bravyi_kitaev_2012, Toloui_2013, peruzzo_variational_2014, wecker_gate_count_2014, poulin_trotter_2014, hastings_improving_2014, mcclean_exploiting_2014, mcclean_exploiting_2014, babbush_chemical_2015, wecker_progress_2015, babbush_exponentially_2016, mcclean_theory_2016, kivlichan_bounding_2017, babbush_exponentially_2018, babbush_low_depth_2018, poulin_quantum_2018,berry_improved_2018, babbush_encoding_2018, berry_qubitization_2019, higgott2019variational, wang2019accelerated, lee_even_2021, von_burg_quantum_2021, huggins_unbiasing_2022,Su2021}. Similarly, but to a lesser degree, quantum algorithms taking into account the specifics of condensed matter applications have also been conceived. These include the development of different flavors of variational quantum eigensolvers (VQE)~\cite{manrique2020momentum,clinton_towards_2022,song_periodic_2022,Yoshioka2022}, the quantum imaginary time evolution algorithm~\cite{motta_determining_2020}, and fault-tolerant algorithms~\cite{babbush_low_depth_2018,kivlichan_improved_2020,campbell_early_2021,kanno_resource_2022, flannigan2022propagation,Su2021} for simulation of model Hamiltonians, such as the Hubbard model, as well as first-principles Hamiltonians. 

Quantum computers can provide a computational advantage over classical computers only for hard classical problems. These include the simulation of so-called strongly correlated systems and, more practically, problems that are not solved with sufficient accuracy using classical methods with low computation cost --- such as Kohn-Sham density functional theory (KS-DFT))~\cite{Hohenberg1964,Kohn1965,Kohn1999} or coupled-cluster theory~\cite{vcivzek1966correlation,shavitt2009many}. Notwithstanding varying definitions and interpretations of ``strong correlation'', and an ongoing debate regarding the extent to which KS-DFT can describe such systems~\cite{cremer2001density,Perdew2021}, the general consensus is that molecular and solid state systems with a large number of localised $d$ or $f$ electrons present a significant challenge for classical simulations. Examples of such systems include transition metal oxides such as NiO and PdO used in heterogeneous catalysis applications. The number of localised sites in such systems is formally infinite as the solids should be simulated at the thermodynamic limit. In practice, one restricts calculations to a periodic finite-sized cell (also referred to as supercell) with {\it ca.}~30--100 unique transition metal atoms; all other atoms in the solid are replicas of those in this computational cell.

The ability to accurately model the electronic structure of materials such as NiO and PdO would no doubt prove extremely useful in the study of heterogeneous catalysis, a field with no shortage of materials that are poorly described by DFT. It is often the case that the interpretation of calculated results (\textit{e.g.}~regarding trends in activity) must be presented with significant caveats regarding the underlying nature of the models used.

In this work, we focus on the calculation of the ground state energy of electrons in materials within the Born-Oppenheimer approximation~\cite{born_zur_1927}. This corresponds to finding the lowest eigenvalue of the electronic Hamiltonian for a fixed position of the nuclei. Two main families of quantum algorithms can perform such calculation: VQE~\cite{peruzzo_variational_2014} and quantum phase estimation (QPE)~\cite{Kitaev1995,cleve1998quantum,NielsenChuang}. While VQE might have its merits in certain use cases, it appears the emerging consensus is that QPE has a superior scaling with the system size~\cite{blunt2022perspective,liu_prospects_2022}.
In order to estimate the eigenvalues of the Hamiltonian with QPE, one has to implement a unitary operator encoding the spectrum of the Hamiltonian. QPE requires deep quantum circuits, and as such it will need to run on error-corrected quantum computers. In such error-corrected implementations one must strive to minimize the number of T gates needed to encode the Hamiltonian, as these gates are the costliest to implement  (see e.g.~\cite{fowler_low_2018}). 
To this date, the most cost-efficient approaches for such encodings are based on the so-called qubitization technique~\cite{low_hamiltonian_2019,poulin_quantum_2018,berry_improved_2018,berry_improved_2018,babbush_encoding_2018,berry_qubitization_2019,von_burg_quantum_2021,lee_even_2021}. Previous work on second quantized Hamiltonians for realistic solid state systems has mainly focused on the Trotterization approach~\cite{babbush_low_depth_2018,kivlichan_improved_2020}. In this work, we adapt the sparse qubitization approach to the simulation of crystalline solids with QPE and estimate the resources required for calculating the ground state energy of crystals in error-corrected quantum computers.

The quantum resources required to simulate a Hamiltonian strongly depend on the single-electron basis sets used to represent electron interactions. For crystalline solids, plane waves (PW) currently appear to be one of the most efficient basis sets both in first and second quantization~\cite{Su2021,babbush_low_depth_2018}. An advantage of using PW basis sets is the sparse representation of the electronic Hamiltonian. This advantage is always exploited in classical computations such as KS-DFT~\cite{Payne1992, Kresse1996prb}. The number of two-body terms in PW representation scales cubically with the size of the basis set. The main disadvantage of such basis sets, however, is that they require a large number of basis functions, especially in all-electron calculations. For crystalline solids one can exploit Bloch functions instead, which are plane waves times a periodic function with the periodicity of the unit cell. In the Bloch representation, the number of terms also scales cubically with the system size, and at the same time such a representation allows using localised atomic orbitals as the periodic constituent of the orbitals. The other commonly used representation in computational condensed matter physics is the Wannier representation, in which orbitals are localized in space~\cite{Marzari1997,Skylaris2002, Marzari2012}. Wannier orbitals can be related to Bloch functions through Fourier transformation, and can be localized using unitary optimization in order to produce maximally localized Wannier functions~\cite{Marzari1997}. When the periodic function in the Bloch representation is a constant, the Wannier representation coincides with the PW dual representation introduced in the context of quantum computing in Ref.~\cite{babbush_low_depth_2018}. At the same time, Wannier orbitals can be spanned in the localised atomic orbital basis which in turn can significantly reduce the size of the basis set for an accurate description of finite band-gap solids. In this work, we investigate Bloch and Wannier representations in the context of qubitized QPE. We note that such basis sets have recently been investigated in the context of the VQE algorithm~\cite{clinton_towards_2022}. 

Quantum computation with qubitization-based QPE requires a large number of gates in a circuit. In order to perform large quantum computations, one has to encode a logical qubit using several physical qubits with a technique known as quantum error correction~\cite{shor_scheme_1995,roffe_quantum_2019}. In order to estimate the total number of physical and logical qubits required for the implementation of quantum algorithms as well as their runtime, we have followed Litinski's approach~\cite{litinski_game_2019}. This scheme operates the surface code~\cite{fowler_surface_2012} with lattice surgery~\cite{horsman_surface_2012,fowler_low_2018}, and compiles logical quantum circuits down to just multi-qubit T gates and multi-qubit measurements---all Clifford gates are commuted past the end of the circuit. In this way, runtime is directly related to T-gate count.

The article is organized as follows. In Sec.~\ref{sec:phys_systems}, we first describe the relevance of modelling bulk materials such as NiO and PdO for applications to heterogeneous catalysis---an area where quantum computation can provide high accuracy results when error-corrected quantum computers become available. In Sec.~\ref{sec:methodology}, the Hamiltonian, basis sets, and quantum algorithms for modelling of crystalline solids are introduced. In Sec.~\ref{sec:results}, we discuss the performance of quantum algorithms and provide quantum resource estimations for several solid state systems. Finally, discussion and conclusions are presented in Sec.~\ref{sec:disc_and_concl}. Detailed logical qubit and Toffoli gate counts of the sparse qubitization are provided in Appendix~\ref{sec:detailed costings}.

\section{Materials and Heterogeneous Catalysts}\label{sec:phys_systems}
Catalysts are used in practically every industrial chemical process, with applications in agriculture, transportation and energy production, among many others. The function of a catalyst to ultimately reduce the energy requirements of a process to make it viable or more efficient means that catalytic processes are a key component for ensuring a sustainable future and reducing human impact on the environment.
%
Transition metal oxide catalysts are essential components for many important industrial processes (such as refining and petrochemistry, fuel cells, hydrogen production, biomass conversion, photocatalysis) where they are used both directly, as the active material (providing the active site), and indirectly, as a support material (commonly as a reducible oxide taking a secondary role in the catalysis).
The overall performance of the solid catalyst depends on many factors, including the particle size, particle shape, crystallinity, chemical composition, and all preparation and activation procedures. High catalytic efficiencies are achieved as the number of active surface sites grows, while the structural flexibility of supported metal catalysts (dynamic structural changes) is key for the catalytic reactivity when we consider that the surface sites repeatedly participate in adsorption/desorption cycles.

The systems considered in this work, nickel oxide (NiO) and palladium oxide (PdO), both form the basis of industrially relevant catalyst materials. In the field of energy and environment, natural gas reforming is the most common process used in industry to produce H$_2$ from fossil fuels, known as methane steam reforming (MSR). Here NiO is reduced to Ni which functions as a high-temperature catalyst. Despite its age and ubiquity, the MSR process still has many technical challenges, for instance around deactivation from carbon whisker formation and stability at high-temperature. Under certain operating conditions, a local oxidizing environment can form within the reactor, leading to the deactivation of Ni due to NiO being present. Thermodynamics can predict the conditions at which this can occur \cite{twigg_catalyst_2018}. However, in general, it is still a challenge to obtain reliable or accurate thermodynamic parameters for strongly correlated oxide materials, especially when they deviate form the bulk limit such as in nanoparticles.

Methane has an estimated greenhouse warming potential (GWP 100) of 27.9 \cite{masson-delmotte_working_2021}, meaning its emissions contribute significantly to global warming and climate change; it is therefore necessary to reduce them wherever possible. Among the many different technologies for methane abatement, methane combustion catalysts based on palladium can be found. Such technologies include after-treatment for combustion of natural gas engines (CNG) and diesel oxidation catalysts (DOC) as well as in mine ventilation systems.

In the above applications, methane is efficiently combusted over palladium (or alloyed) oxide catalyst to produce H$_2$O and CO$_2$, with activity in this process influenced by a number of factors. A technical target in practical catalysis is to reduce the temperature at which this occurs, allowing for a lower operating temperature and more efficient handling of emissions. Partial oxidation can sometimes occur, and may indeed be desirable in the development of processes to produce precursors for more complex chemicals. The ability to simulate accurately not only the activity but also the selectivity, which is a measure of a catalyst's ability to promote the formation of the desired product(s) over other possibilities, is crucial to the prediction of new catalysts.

\begin{figure}
 \centering
	   \includegraphics[width=0.6\textwidth]{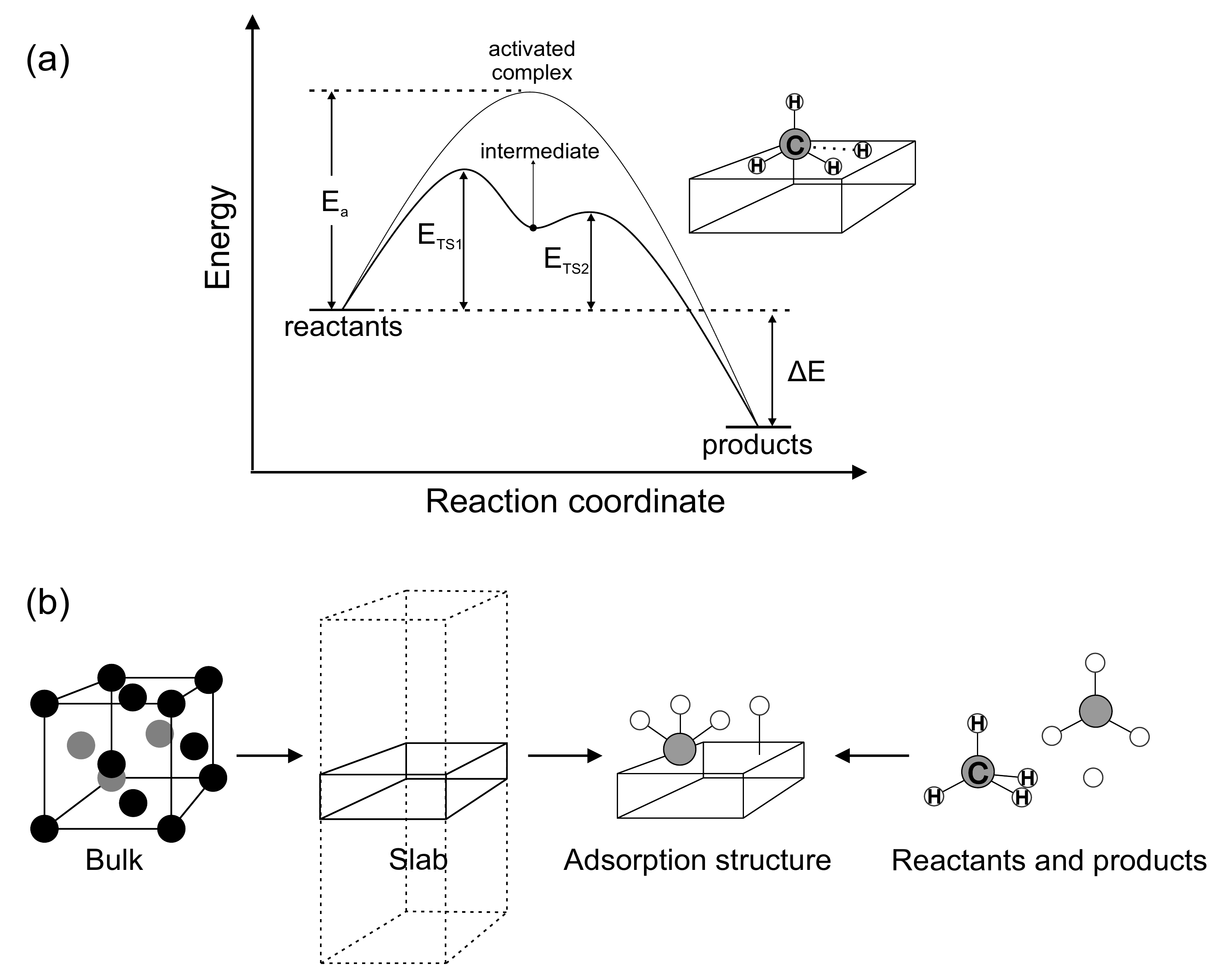}
 \caption{(a) Schematic of a reaction pathway with and without a catalyst, and (b) diagram of the necessary calculation steps required to determine these properties. Relative energies between the reactants, products and intermediates (if they are known) can be calculated via density functional theory or other computational approaches. Reaction barriers for a catalyzed reaction can be approximated via adsorption energy calculations in many cases \cite{cheng_bronstedevanspolanyi_2008}, or by transition state searches, which require the calculation of forces as well as the energy.}
 \label{fig:catalyst_schematic}
\end{figure}

Figure~\ref{fig:catalyst_schematic}(a) shows a schematic of a typical catalyzed reaction. The presence of a catalyst provides additional reaction coordinates, or reaction intermediates, with their own activation energies ($E_{TS1}$, $E_{TS2}$). For an effective catalyst, these energies are necessarily lower than the uncatalyzed activation energy $E_a$.
In the case of heterogeneous catalysts, reaction intermediates are typically adsorption steps, where one or more of the reactants binds to a surface site of the catalyst. Depending on the complexity of the reaction mechanisms, there may be a large number of these intermediates as well as branches and side reactions that must be considered when studying a reaction in order to determine the key step(s). It is often necessary to find these steps, which govern the activity and/or selectivity of a catalyst, as in doing so, the problem is reduced to fewer dimensions and descriptors that facilitate a more rapid study. For example, in a kinetic analysis the largest activation energy is usually of most interest, as this will be the rate determining step. Whilst this knowledge may be well established in well known reactions, it can be necessary to perform many calculations in more novel applications. Furthermore, whilst the accuracy of current computational approaches may be good enough to predict trends in similar systems, obtaining chemical accuracy and absolute values for detailed kinetic studies remains a challenge. Figure~\ref{fig:catalyst_schematic}(b) shows a typical set of model systems that would be used to estimate the energetics of a heterogeneous catalytic process.

When running simulations of a catalyst, consideration needs to be made of the question at hand and the level of accuracy that is needed. Broadly speaking, we are interested in activity, selectivity and stability. When simulating activity, we often need a kinetic model which can provide rates or turn-over frequencies. If we are interested in screening for materials, it is often sufficient to correlate these rates with descriptors \cite{toulhoat_kinetic_2003}.

For example, following the Sabatier principle \cite{medford_sabatier_2015}, which is employed primarily for materials screening, calculating the (heterogeneous) catalytic activity of a material is performed by determining the binding strengths of the reactants, products and any important intermediates of a given reaction with the surface of that material. These binding strengths can be determined from energy calculations using a wide variety of models, each with their own trade-offs between accuracy, transferability and computational cost.

However, if we are interested in predicting reactor performance or process conditions then we need significantly greater precision in the simulated parameters. Likewise, simulating the often subtle differences in competing reactions (which result in different products) typically requires greater accuracy in calculations to predict selectivity.

Whilst the questions of activity and selectivity are crucial for a material's function as a catalyst, when looking for a technical solution, the question of stability becomes critical. Catalysts often need to operate over many years under harsh conditions (high temperature, pressure, contaminated conditions and, in the case of electrocatalysis, high potentials and corrosive environments). The simulation of stability introduces a whole range of other problems; for instance, predicting morphological changes and thermal degradation of a catalyst requires a large number of calculations, often of large model systems, to allow sintering of nanoparticles or ceramic supports to be conducted. Material complexity (\textit{e.g.} simulation of realistic metal/ceramic interfaces), bridging time and length scales where accurate atomic-scale materials properties can be fed into multi-scale models, are all open challenges in this area.

DFT is one of the most successful and widely used models for calculating the energies of molecular and solid state systems relevant to industrial processes. It is an \textit{ab initio} method that uses functionals of the electronic density to calculate energy rather than attempt to deal directly with the many-body wavefunction. In KS-DFT, the electronic density is constructed using a fictitious set of non-interacting single electron wavefunctions and approximating an unknown correction term. This term, known as the exchange-correlation (XC) functional, includes exchange and correlation effects as well as discrepancy between the real and non-interacting kinetic energy. There are many choices, though all of them approximated, for its form.

Ultimately, it is the use of single-particle wavefunctions in DFT that leads to some of its most prominent shortcomings. In the case of NiO, and indeed most transition metal oxides, the strong electron-electron interactions of the d-electrons in these materials is poorly described by approximate KS-DFT, leading to over-delocalisation of these bands (and to the prediction of more metallic electronic structures than the reality). A Hubbard U \cite{kulik_perspective_2015} correction can be used alongside local density approximation (LDA) and generalised gradient approximation (GGA) XC functionals to mitigate this issue in some cases, although it is overly empirical in nature. While the use of hybrid XC functionals such as PBE0 \cite{adamo_toward_1999} can sometimes perform better \cite{mandal_systematic_2019}, due to the inclusion of Hartree-Fock exact exchange, the fraction of exact exchange to use can be varied (depending on the XC functional used), which again leads to empirical fitting. Hybrid functionals are also incomplete (and incorrect) in their description of the electronic structure, and are by no means a guaranteed improvement over GGA functionals in their prediction of transition metal oxide properties \cite{coulter_limitations_2013}.

To model the bulk properties of materials effectively, the use of periodic boundary conditions (PBCs) is required, allowing for a simulation box to include only the primitive unit cell in highly ordered systems. Even in disordered systems, periodicity is still imposed (on a larger unit cell), as the approximation still provides more representative models than any non-periodic alternative, without extending the system far beyond practical limits.

The study of heterogeneous catalysis primarily concerns the properties of surfaces, so slab models are often used. These are also periodic, albeit in 2 dimensions rather than 3. Bulk calculations are also required in order to determine the surface energies of the facets of a material, which, for example, allow for the prediction of the expected shape of nanoparticles, as well as which facets are most predominant and relevant for catalysis. The stability of a material is another important aspect that can be predicted by energy calculations on bulk systems.

\section{Methodology}\label{sec:methodology}

\subsection{Hamiltonian for Periodic Systems}\label{sec:HamPeriodicSys}
The Hamiltonian of interacting electrons in the Born-Oppenheimer approximation can be written as follows:
\begin{equation} \label{eq: Ham-Field}
    \hat H = \hat H^{(0)} + \hat H^{(1)} + \hat H^{(2)},
\end{equation}
where $\hat H^{(0)}$ is a constant term describing nuclear repulsion, $\hat H^{(1)}$ and $\hat H^{(2)}$ are one-body and two-body terms, respectively~\cite[p. 32]{Stefanucci2013}:
\begin{equation}
    \hat H^{(1)} = \IINT{} \hat \psi^{\dagger}(\tx) h(\tx, \tx') \hat \psi(\tx')
\end{equation}
\begin{equation}
    \hat H^{(2)} = \frac{1}{2}\IINT{} \hat  \psi^{\dagger}(\tx) \hat \psi^{\dagger}(\tx') g(\tx,\tx') \hat \psi(\tx') \hat \psi(\tx),
\end{equation}
$\tx$ denotes position and spin, $(\tr, \sigma)$, of an electron and the integration domain is over the volume of the macroscopic crystal, $V$. In this work, we do not consider the external magnetic field or spin-orbit coupling and therefore, the one- and two-body kernels are diagonal w.r.t. spin degrees of freedom. The spatial part of one-body kernel is:
\begin{equation}
h(\tr) 
= -\frac{1}{2}\nabla^2 + U(\tr) 
\end{equation}
where $U(\tr)$ is the nuclei potential
\begin{equation}
U(\tr) = \sum_{a\in V} \frac{Z_{a}}{|\tr - {\bf P}_a|},
\end{equation}
$Z_{a}$ and ${\bf P}_a$ are the nuclear charge and position of nucleus $a$. 
The spatial part of two-body kernel is
\begin{equation}
g(\tr, \tr') = g(|\tr - \tr'|) = \frac{1}{|\tr - \tr'|}
\end{equation}
We assume Born-von-K\'arm\'an periodic boundary conditions at the boundaries of the macroscopic crystal which is defined by the vectors ${\bf L}_1, {\bf L}_2, {\bf L}_3$:
\begin{align}
    A(\tr + {\bf L}_\alpha) & = A(\tr), \quad A = \hat \psi, h, U, g \quad \alpha = 1, 2, 3
\end{align}
In this case, the external potential and two-body kernel are defined in terms of their Fourier series:
\begin{align}\label{eq: external_fourier_full}
    U(\tr) & = \sum_{a \in V} \sum_{{\bf K}} \frac{4\pi Z_a e^{-i {\bf K}{\bf P}_a}}{{V \bf K}^2} e^{i {\bf K} \tr} \\
    g(\tr) & = \sum_{{\bf K}} \frac{4\pi}{V{\bf K}^2} e^{i {\bf K} \tr} 
\end{align}
where ${\bf K}$ satisfies:
\begin{equation}
    {\bf K}{\bf L}_\alpha = 2\pi m_\alpha, \quad m_\alpha \in \mathbb{Z}, \quad \alpha=1, 2, 3
\end{equation}
Crystalline solids consist of unit cells and each unit cell $V_{uc}$ is defined by translation lattice vectors, $\ta_1, \ta_2, \ta_3$.  Each unit cell can be labeled with $\tR$ indicating a node of the Bravais lattice:
\begin{equation}
    \tR = A^{T} \tn; \quad \tn = (n_1, n_2, n_3)^T,\, n_\alpha \in \mathbb{Z}; \quad A = [\ta_1, \ta_2, \ta_3]^T 
\end{equation}
Let $N_\alpha \ge 1$ be the number of unit cells along $\ta_\alpha$, $n_\alpha = 0, 1, .., N_\alpha-1$ and thus, the total number of unit cells which spans the whole finite macroscopic crystal is $N=N_1 N_2 N_3$. 
We also introduce the reciprocal lattice which is defined as:
\begin{equation}
    \tG = B \tn; \quad \tn = (n_1, n_2, n_3)^T,\, n_\alpha \in \mathbb{Z}; \quad B = [\tb_1,\tb_2,\tb_3]=2\pi A^{-1}
\end{equation}
The vectors $\tb_1,\tb_2,\tb_3$ and $\ta_1, \ta_2, \ta_3$ satisfy the following relations:
\begin{equation}
    \ta_\alpha \tb_\beta = 2\pi \delta_{\alpha\beta}
\end{equation}

In the case of crystalline solids, the external potential can also be rewritten in terms of reciprocal lattice vectors, because it is has periodicity of the lattice: 
\begin{align}
    U(\tr) & = \sum_{a \in V_{uc}} \sum_{{\bf G}} \frac{4\pi Z_a e^{-i {\bf G}{\bf P}_a}}{{V_{uc} \bf G}^2} e^{i {\bf G} \tr} 
\end{align}
This is similar to Eq.~\eqref{eq: external_fourier_full} but written for unit-cell periodicity instead of periodicity within macroscopic crystal.

In order to perform practical calculations, one can choose a single-particle basis which is suitable for the problem of interest:
\begin{equation}
    \hat \psi(\tx) = \sum_{p, \sigma} \psi_{p}(\tr) \aps,
\end{equation}
where $p$ can be a set of numbers describing a single particle state such as wave-vector index and band index, for example. 
The Hamiltonian in the new basis set can be written as:
\begin{equation} \label{eq: Ham-Ferm-Chem}
    \hat H = H^{(0)} + \sum_{pq\sigma}\left(h_{pq} - \frac{1}{2} \sum_{r} g_{prrq}\right) \adps \aqs + \frac{1}{2} \sum_{pqrs; \sigma,\tau} g_{pqrs} \adps \aqs \adrt \ast
\end{equation}
where two-body matrix elements (often referred to as the electron-repulsion integrals or just Coulomb integrals) are:
\begin{equation}
    g_{pqrs} = \IINTr{V} \psi^{*}_{p}(\tr) \psi_{q}(\tr) \psi^{*}_{r}(\tr') \psi_{s}(\tr') g(|\tr - \tr'|)
\end{equation}
From this definition, $g_{pqrs}$ obeys the 4-fold symmetry relations:
\begin{equation} \label{eq: g_4fold}
    g_{pqrs} = g_{rspq} = g_{qpsr}^{*} = g_{srqp}^{*} 
\end{equation}
If coefficients are real then $g_{pqrs}$ obeys the 8-fold symmetry relations:
\begin{equation}
    g_{pqrs} = g_{pqsr} = g_{qprs} = g_{qpsr} = g_{rspq} = g_{srpq} = g_{rsqp} = g_{srqp} \label{eq: g_8fold}
\end{equation}

The set of orbitals $\{ \psi_{p}(\tr) \}$ should be an orthonormal basis set which satisfies the periodic boundary conditions. Below we describe two basis sets which are commonly used in computational condensed matter physics. Further in the text, the number of spatial orbitals per unit cell is denoted as $M$ (the number of bands) while the total number of spatial orbitals in the crystal is $P=MN$.

\subsubsection{Bloch Functions as a Basis Set}
Bloch functions are the solution of a mean-field problem
in a periodic potential. The Bloch functions can be written as follows~\cite[p.167]{KittelISSP}:
\begin{equation}
    \psi_{\tk j}(\tr) = e^{i\tk\tr}u_{\tk j}(\tr) \label{eq:Bloch functions}
\end{equation}
where $u_{\tk}(\tr)$ is periodic function with periodicity of the unit cell, $u_{\tk}(\tr) = u_{\tk}(\tr + \tR)$, $j$ 
is the band index, $0 \leq j \le M$ and $\tk = (k_1, k_2, k_3)^{T}$ is the wave vector belonging to the first Brillouin zone which can be defined as:
\begin{equation}
    k_\alpha = \sum_{\beta=1,2,3} B_{\alpha\beta} \frac{g_\beta}{N_\beta}, 
\end{equation}
where $g_\beta$ is an integer such that
\begin{equation}\label{eq: gi_range}
    -\frac{N_\alpha}{2} < g_\alpha \leq \frac{N_\alpha}{2}
\end{equation}
The larger the size of the macroscopic crystal, the larger the number of $k$-points in the Brillouin zone as can be seen from \eqref{eq: gi_range}. 
Using Bloch functions as the basis set
\begin{equation}
    \hat \psi(\tx) = \sum_{\tk j} \psi_{\tk j}(\tr) \hc_{\tk j \sigma}
\end{equation}
the Hamiltonian can be written as: 
\begin{equation}
    \hat H =
    H^{(0)} + 
    \sum_{\tk ij \sigma} 
    \tilde{h}_{\tk ij} \hc^{\dagger}_{\tk i \sigma} \hc_{\tk j \sigma} 
    + 
    \frac{1}{2N} \sum_{\sigma, \sigma'}\sum_{ijlm} 
    \sum_{\tk\tq\tk'\tq'}
    \left(\sum_{\tG} \delta_{\tq+\tq'-\tk-\tk', \tG} \right)
    \tilde{g}_{\tk i, \tq j, \tk'l, \tq' m} 
    \hc^{\dagger}_{\tk i \sigma} \hc^{\dagger}_{\tk' l \sigma'} \hc_{\tq' m \sigma'} \hc_{\tq j \sigma} 
    \label{eq:Bloch hamiltonian}
\end{equation}
where we used the convention that the Bloch functions are normalized in the unit cell.
Due to $\sum_{\tG} \delta_{\tq+\tq'-\tk-\tk', \tG}$, the number of two-body terms scales as $O(N^3)$, the same as in the plane-wave basis set~\cite{babbush_low_depth_2018}. One- and two-body matrix elements are defined as:
\begin{equation}
  \tilde{h}_{\tk ij} = \INTr {V_{uc}} \psi^{*}_{\tk i}(\tr) h(\tr) \psi_{\tk' j}(\tr)
\end{equation}
and
\begin{equation}
    \tilde{g}_{\tk i, \tq j, \tk'l, \tq' m} = 
    \INTr{V_{uc}} 
    \psi_{\tk i}^{*}(\tr)\psi_{\tq j}(\tr) 
    \INTr{V_{uc}}' 
    \psi_{\tk' l}^{*}(\tr')\psi_{\tq' m}(\tr') 
    g_{\tq'-\tk'}(\tr-\tr'),
\end{equation}
where 
\begin{align}
    g_{\tq-\tk}(\tr-\tr') = 
    \sum_{\tG} \frac{4\pi}{(\tq - \tk + \tG)^2 V_{uc}} e^{i(\tq - \tk + \tG) (\tr-\tr')}
\end{align}
and $V_{uc}$ is the volume of the unit cell. Coefficients in the Hamiltonian are usually complex and the two-body terms obey 4-fold symmetry~\eqref{eq: g_4fold}. The composite index from Sec.~\ref{sec:HamPeriodicSys} indicates both band and wave vector, $p=(\tk, i)$, $q=(\tq, j)$, $r=(\tk', l)$, $s=(\tq', m)$.

\subsubsection{Wannier Functions as a Basis Set}

Wannier functions are the set of localized orbitals which obey the translational symmetry of the crystal:
\begin{equation} \label{eq: Wannier symmetry}
    \phi_{\tR j}(\tr - \tR') = \phi_{\tR + \tR' j}(\tr)
\end{equation}
Such localized orbitals can be obtained by carrying out a localization procedure in the supercell or applying a Fourier transformation to the Bloch orbitals~\cite{Marzari2012}:
\begin{equation}
    \phi_{\tR j}(\tr) = \frac{1}{N} \sum_{\tk} e^{-i\tk \tR} \sum_{m}\psi_{\tk m}(\tr) U_{mj}^{(\tk)},
\label{eq:Wannier functions}
\end{equation}
where unitary matrices can be chosen according to localization criteria such as, for example, Foster-Boys~\cite{Foster1960} (Maximally Localized Wannier orbitals~\cite{Marzari1997}) or Pipek-Mezey~\cite{Pipek1989,lehtola_pipek_2012,Jonsson2017}. Contrary to Bloch functions, these functions can be chosen to be real valued~\cite{Marzari2012}.
In the Wannier basis set
\begin{equation}
    \hat \psi(\tx) = \sum_{\tR j} \phi_{\tR j}(\tr) \ha_{\tR j \sigma}
\end{equation}
the Hamiltonian is
\begin{equation}
    \hat H = H^{(0)} + 
    \sum_{\tR \tR' ij \sigma} 
    h_{\tR p, \tR'q} \ha^{\dagger}_{\tR i \sigma} \ha_{\tR j \sigma} + 
    \frac{1}{2} 
    \sum_{\sigma, \sigma'}
    \sum_{ijlm} 
    \sum_{\tR\tT\tR'\tT'}
    g_{\tR i, \tT j, \tR'l, \tT' m} 
    \ha^{\dagger}_{\tR i \sigma} 
    \ha^{\dagger}_{\tR' l \sigma'} 
    \ha_{\tT' m \sigma'} 
    \ha_{\tT j \sigma} 
\end{equation}
with matrix elements:
\begin{equation}
    h_{\tR i, \tR'j} =
  \INTr {V} \phi^{*}_{\tR i}(\tr) h(\tr) \phi_{\tR' j}(\tr)
\end{equation}
and
\begin{equation}
    g_{\tR i, \tT j, \tR'l, \tT' m} =
    \INTr{V} 
    \phi_{\tR i}^{*}(\tr)\phi_{\tT j}(\tr) 
    \INTr{V}' 
    \phi_{\tR' l}^{*}(\tr')\phi_{\tT' m}(\tr') 
    g(\tr-\tr')
\end{equation}
which satisfy the following relations:
\begin{align}\label{eq: translational_symmetry_1}
    & h_{\tR i, \tR'j} = h_{0 i, \tR' - \tR j} \\
    & v_{\tR i \tT j \tR'l \tT'm} = v_{0 i, \tT - \tR j, \tR'-\tR l, \tT'-\tR m} \label{eq: translational_symmetry_2}
\end{align}
This is due to the fact that Wannier functions obey Eq.~\eqref{eq: Wannier symmetry}.
In this paper we don't construct Wannier functions from Bloch orbitals but rather choose natural atomic orbitals in the supercell calculations~(see Section~\ref{subsec: classical_computational_details} for computational details). Since the two-body term is real, it satisfies Eq.~\eqref{eq: g_8fold}. The composite index from Sec.~\ref{sec:HamPeriodicSys} indicates both band and unit cell indices, $p=(\tR, j)$, $q=(\tT, j)$, $r=(\tR', l)$, $s=(\tT', m)$.

\subsubsection{Majorana Representation}\label{sec:majorana representation}
Majorana operators represent a convenient choice for working with quantum computing algorithms. The reason is that each Majorana operator is Hermitian and can be mapped onto one Pauli string using a qubit representation and, as a result, any unique product of Majorana operators is one Pauli string. The actual qubit representation depends on the choice of transformation such as the Jordan-Wigner~\cite{Jordan1928,ortiz_quantum_2001} or Bravyi-Kitaev~\cite{Bravyi2002,seeley_bravyi_kitaev_2012}. However, some properties of the Hamiltonian which do not depend on the choice of qubit mapping can conveniently be obtained in Majorana representation. We will use this representation in order to generalize the sparse qubitization approach on Hamiltonians with complex coefficients. Majorana operators are defined as:
\begin{align}
    \pso & = \aps + \adps\\
    \psn & = -i(\aps - \adps)
\end{align}
with an additional binary index specifying the Majorana type.

They satisfy the following anti-commutation relations:
\begin{equation}
\{\psk, \qtj \} = 2 \delta_{pq} \delta_{\sigma\tau} \delta_{kj}
\end{equation}

Following Refs.~\cite{von_burg_quantum_2021,Koridon2021}, the constant, one-body and two-body terms of the Hamiltonian~\eqref{eq: Ham-Ferm-Chem} in Majorana representation can be written as follows:
\begin{align}\label{eq: Maj-Ham-H0}
    \tilde{H}_0 = H_0 + 
    \sum_{p} h_{pp} + \frac{1}{2}\sum_{pq} \left(g_{ppqq} - \frac{1}{2} g_{pqqp}\right)
\end{align}
\begin{align}
    \hat{\tilde{H}}^{(1)} & =
    \frac{i}{2}
    \sum_{p\le q} 
    {\rm Re}\left(h_{pq} + \frac{1}{2}\sum_{r} \left(2 g_{pqrr}  - g_{prrq}\right) \right)
    \left(\frac{1}{2}\right)^{\delta_{pq}}
    \sum_{\sigma;j\neq k} (-1)^j
    \psj\qsk \label{eq:1body re}
    \\& + 
    \frac{i}{2} \sum_{p<q}  
    {\rm Im}\left(h_{pq} + \frac{1}{2}\sum_{r}
    \left(2g_{pqrr} - g_{prrq}\right)\right) 
    \sum_{\sigma; j}
    \psj\qsj \label{eq:1body im}
\end{align}
\begin{align}
    \hat{\tilde{H}}^{(2)} & =
    \frac{1}{4}\sum_{\substack{p\le q, r\le s \\ (p,q)\le (r,s)}} 
    B_{pqrs} \left(\frac{1}{2}\right)^{\delta_{pq}+\delta_{rs}+\delta_{(p,q),(r,s)}}
    \sum_{\sigma\neq\tau, j\neq k, l\neq m}
    \psj\qsk \rtl \stm (-1)^{j+m} \label{eq:2body B}\\&
    +
    \label{eq: F}
    \frac{1}{4}\sum_{\substack{p < q, r < s \\ (p,q)\le(r,s)}}
    F_{pqrs} 
    \left(\frac{1}{2}\right)^{\delta_{(p,q),(r,s)}}
    \sum_{\sigma\neq\tau; j,k}
    \psj\qsj\rtk\stk  (-1)^{j+k}\\&
    + 
    \label{eq: C}
    \frac{1}{4}\sum_{p\le q, r<s} 
    (-1)C_{pqrs} \left(\frac{1}{2}\right)^{\delta_{pq}} 
    \sum_{\sigma\neq\tau; j\neq k; l}
    \psj\qsk \rtl \stl (-1)^{j} \\&
    +
    \label{eq: 2body BBF}
    \frac{1}{4}\sum_{\substack{p < q, r < s \\ (p,q)\le(r,s)}} 
    \left(F_{pqrs} + B_{prqs} - B_{psqr}\right)
    \left(\frac{1}{2}\right)^{\delta_{(p,q),(r,s)}} 
    \sum_{\sigma, j\neq k}
    \psj\qsj\rsk\ssk  \\& 
    +
    \label{eq: F-F-F}
    \frac{1}{4}
    \sum_{p < q < r < s}(F_{pqrs}  - F_{prqs} + F_{psqr})
    \sum_{\sigma,j}
    \psj\qsj\rsj\ssj  \\&
    +
    \label{eq: C-C-C}
    \frac{1}{4} \sum_{p, q < r < s} \left(C_{pqrs} - C_{prqs} + C_{psqr}\right)
    \sum_{\sigma, j\neq k}
    \psj\qsk\rsk\ssk (-1)^j 
\end{align}
where the tensors
\begin{align}
    \label{eq:Bpqrs}
    & B_{pqrs} = \frac{{\rm Re}(g_{pqrs} + g_{pqsr})}{2} = \frac{g_{pqrs} + g_{qpsr} + g_{pqsr} + g_{qprs}}{4} \\
    \label{eq:Cpqrs}
    & C_{pqrs} = \frac{{\rm Im}(g_{pqrs} - g_{pqsr})}{2} = \frac{g_{pqrs} - g_{qpsr} - g_{pqsr} + g_{qprs}}{4i} \\
    \label{eq:Fpqrs}
    & F_{pqrs} = \frac{{\rm Re}(g_{pqrs} - g_{pqsr})}{2} = \frac{g_{pqrs} + g_{qpsr} - g_{pqsr} - g_{qprs}}{4}
\end{align}
$B_{pqrs}$ is symmetric w.r.t.~$(p, q)$ and symmetric w.r.t.~$(r, s)$,
$C_{pqrs}$ is symmetric w.r.t.~$(p, q)$ and antisymmetric w.r.t.~$(r, s)$,
$F_{pqrs}$ is antisymmetric w.r.t.~$(p, q)$ and antisymmetric w.r.t.~$(r, s)$,
$B$ and $F$ are symmetric w.r.t.~interchange of pairs $(p, q)$, $(r, s)$, while $C_{pqrs}$ is not.  \\
The reverse relation is 
\begin{equation}
    g_{pqrs} = B_{pqrs} + F_{pqrs} + i(C_{rspq} +  C_{pqrs}).
\end{equation}
This representation of the Hamiltonian is valid for both Bloch and Wannier orbitals and the only differences are indices labeling states and the value of coefficients.
For real-valued Wannier functions, only the $B_{pqrs}$ tensor is non-zero  while for complex-valued orbitals, such as Bloch functions, all tensors need to be taken into account as coefficients of the Hamiltonian in Eq.~\eqref{eq: Ham-Ferm-Chem} can be complex. 
The Hamiltonian in Majorana representation~\eqref{eq: Maj-Ham-H0}--\eqref{eq: C-C-C} provides a decomposition in a linear combination of unitaries (LCU~\cite{Childs2012}) and the entry-wise L$_1$ norm of such a Hamiltonian, which is the same in any qubit representation~\cite{Koridon2021}, can be written as:
\begin{equation}
    \alpha = \lambda_0 + \lambda = \lambda_0 + \lambda_1 + \lambda_2
\end{equation}
where
\begin{equation}
    \lambda_0 =  \left\vert H_0 + \sum_{p} h_{pp} + \frac{1}{4}\sum_{pr} \left(2 g_{pprr} - g_{prrp}\right) \right\vert
    \label{eq:lambda}
\end{equation}
\begin{align}
    \lambda_1 & = \sum_{pq} \left\vert {\rm Re}\left(h_{pq} + \frac{1}{2}\sum_{r} \left(2 g_{pqrr}  - g_{prrq}\right) \right) \right\vert 
                + 2 \sum_{p<q} \left\vert {\rm Im}\left(h_{pq} + \frac{1}{2}\sum_{r} \left(2g_{pqrr} - g_{prrq}\right)\right) \right\vert 
\end{align}
\begin{align}
    \lambda_2 & = \frac{1}{4}\sum_{pqrs} \left\vert 
    B_{pqrs}
    \right\vert 
    +
    \sum_{p < q, r < s} \left\vert 
    F_{pqrs}
    \right\vert + 
        \sum_{p, q, r < s} \left\vert 
    C_{pqrs}
    \right\vert
    +
     \frac{1}{2}\sum_{p<q, r<s} \left\vert 
     F_{pqrs} + B_{prqs} - B_{psqr}
    \right\vert  + \\&
    +
    \sum_{p < q < r < s} \left\vert 
    F_{pqrs} - F_{prqs} + F_{psqr}
     \right\vert 
     +
     \sum_{p, q < r < s} \left\vert 
    C_{pqrs} - C_{prqs} + C_{psqr}
     \right\vert
\end{align}

In the context of quantum computation, the magnitude of $\lambda = \lambda_1 + \lambda_2$ defines the number of repetitions of controlled-unitary in QPE as discussed below.

\subsection{Quantum Algorithms}

QPE~\cite{Kitaev1995,NielsenChuang} allows to determine the phase of a unitary operator which is implemented in the quantum circuit. 
Originally, Hamiltonian simulation with the choice of unitary
\begin{equation}
    U = e^{iHt} \label{eq:trotter U}
\end{equation}
was used, as the time evolution operator can be implemented with Trotterization~\cite{suzuki1991general, childs2018toward} and its phases $E_jt$ are directly related to the system's energies $E_j$. Other methods for Hamiltonian simulation have since been developed, including Taylor series~\cite{babbush_exponentially_2016,babbush_exponentially_2018,babbush_low_depth_2018} and randomised methods~\cite{campbell_random_2019,kivlichan_phase_2019,wan_randomized_2022}.

Instead of the time evolution operator \eqref{eq:trotter U}, one can apply QPE to a unitary \emph{walk operator} $W$ with eigenvalues $e^{\pm i\arccos(E_j/\lambda)}$, from whose phases the Hamiltonian energies can readily be retrieved. This is the approach of more recent qubitisation methods~\cite{poulin_quantum_2018,berry_improved_2018}. The normalization factor $\lambda$ is the $L_1$ norm of Hamiltonian's coefficients in an LCU decomposition, and ensures that all energies correspond to phases between $0$ and $2\pi$. The walk operator is constructed from a reflection along with a \emph{block encoding} of $H/\lambda$.
The starting point for a block encoding is an LCU decomposition of the Hamiltonian
\begin{equation}
    H = \sum_{i=1}^d w_i H_i,\ w_i\in \mathbb{R}^+
    \label{eq:LCU}
\end{equation}
into simpler unitary operators $H_i$ that can be readily implemented on a quantum computer.
In our case, the LCU decomposition of the Hamiltonian is given in section~\ref{sec:majorana representation}, and the $H_i$ consist of strings of up to four Majorana fermions that can be implemented in the Jordan-Wigner representation with a ranged operation \cite{babbush_encoding_2018}.
The two operators
\begin{equation}
    \operatorname{SELECT}\ket{i}\ket{\psi} = \ket{i}H_i\ket{\psi}\ \text{and}\ \operatorname{PREPARE}\ket{0} = \sum_{i=1}^d \sqrt{\frac{{w_i}}{\lambda}}\ket{i}\ \text{with}\ \lambda = \sum_{i=1}^d |w_i|
\end{equation}
facilitate a block encoding of $H/\lambda$ because
\begin{equation}
    \bra{0}(\operatorname{UNPREPARE}\otimes\mathbbm{1})\operatorname{SELECT}(\operatorname{PREPARE}\otimes\mathbbm{1})\ket{0} = H/\lambda
\end{equation}
with $\operatorname{UNPREPARE} = \operatorname{PREPARE}^{-1}$.

The leading order cost of qubitization algorithms is typically proportional to $O(\lambda\sqrt{d}/\epsilon)$.
A multiplicative $\lambda$ is needed to maintain a fixed energy accuracy $\epsilon$ in phase estimation despite the normalization of the Hamiltonian by $\lambda$. Meanwhile, the $\sqrt{d}$ stems from data loading. The values of the $d$ coefficients of the LCU $\omega_i$ can be loaded with Toffoli cost of order $O(\sqrt{d})$ with a select-swap network~\cite{Low2018Trading} (also dubbed QROAM~\cite{berry_qubitization_2019}) at the expense of $O(\sqrt{d})$ ancilla qubits.

Various implementations of the qubitization walk operator~\cite{babbush_encoding_2018,von_burg_quantum_2021,lee_even_2021,berry_qubitization_2019} factorize the
Hamiltonian to find alternative LCUs with lower $d$ and/or $\lambda$. They also show bespoke
implementations of PREPARE and SELECT matching the choice of LCU.
We base our method on the sparse qubitization method \cite{berry_qubitization_2019,lee_even_2021}, which does
not factorize the Hamiltonian, but instead exploits sparsity. In the next section, we will
give an overview of that method. Then we will describe how we have adapted and extended the method to deal
with Bloch and Wannier basis sets. We focus on the asymptotically dominant costs and confine detailed Toffoli and qubit number costings to Appendix~\ref{sec:detailed costings}. The number of T gates is 4 times the number of Toffoli gates~\cite{Gidney2018Halving,babbush_encoding_2018}.

\subsubsection{Overview of Sparse Qubitization}
\label{sec:original sparse qubitisation}
\begin{figure}[H]
\centering
\includegraphics[width=0.9\textwidth]{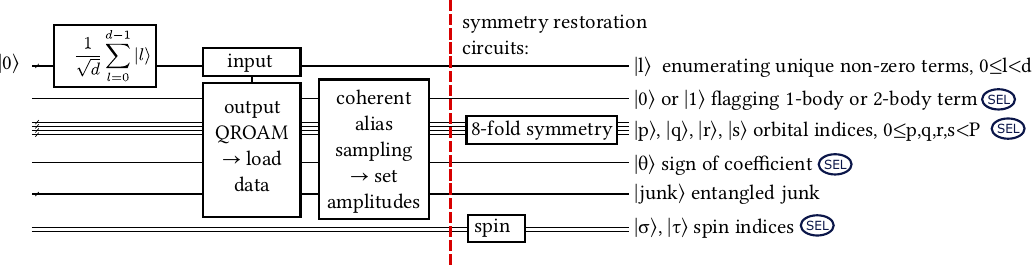}
\caption{PREPARE operator for the original sparse qubitization method \cite{berry_qubitization_2019,lee_even_2021}, outlined in Sec.~\ref{sec:original sparse qubitisation}. It prepares a superposition of product states describing the Hamiltonian's LCU: The amplitudes in the superposition are the (square root modulus of the) LCU's coefficients $\propto V_{pqrs}$, whilst the corresponding product state specifies the unitary operator $Q_{pq\sigma}Q_{rs\tau}$. The state is entangled to an additional junk register, such as the additional qubits needed for coherent alias sampling. Only the qubits indicated with~\includegraphics[height=2ex]{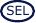}
are used as controls by the SELECT operator. At the vertical dashed line (red), the prepared state describes just one symmetry sector w.r.t.~spin symmetry and 8-fold symmetry. This reduces the amount of data to be loaded, which is the most expensive step in the circuit. The superposition is then enlarged to more product states covering all symmetry sectors with the symmetry restoration circuits.
Note that ancilla qubits (such as in the QROAM or 8-fold symmetry restoration circuit) are not depicted. Despite becoming entangled with the other qubits, like $\ket{\text{junk}}$, they do not affect the rest of the circuit.}
\label{fig:prepare original}
\end{figure}

In this section we outline the original sparse qubitization algorithm which has been used for simulation of real Hamiltonians which satisfy 8-fold symmetry relations~\eqref{eq: g_8fold}. It was first developed in \cite{berry_qubitization_2019} and improved in \cite{lee_even_2021}, Appendix A, on which we base our exposition. Our modifications to this algorithm are presented in the Secs.~\ref{sec:Bloch sparse qubitisation}~and~\ref{sec:Wannier sparse qubitisation}. For simplicity of this summary, we focus on the Hamiltonian's two-body terms, of which there are significantly more than one-body terms.
The main insight the sparse qubitization algorithm \cite{berry_qubitization_2019,lee_even_2021} uses is that the Hamiltonian's LCU decomposition into unitaries $Q$
\begin{equation}
H = \frac{1}{8}\sum_{pqrs,\sigma,\tau} \label{eq:sparse LCU} V_{pqrs} Q_{pq\sigma}Q_{rs\tau}
\end{equation}
is very sparse with respect to the orbital indices, $(p,q,r,s)$, and spin indices, $(\sigma,\tau)$, especially if small coefficients $V_{pqrs}$ are approximated to zero. Thus one can save on quantum resources required for the data loading: Instead of loading the coefficients for all values of $p,q,r,s,\sigma,\tau$, only a unique set of non-zero coefficients is loaded onto the quantum computer and the rest of the Hamiltonian can be restored using symmetry restoration circuits. Let these non-zero coefficients be indexed by $l=1\ldots d$ in an arbitrary way, such that we must load only $d$ data items. 
While the scaling of $d$ with the total number of spatial orbitals, $P$, is still expected to be the same as the full number of electron repulsion integrals,
\begin{equation}
    d \sim O(P^4),
    \label{eq:original d}
\end{equation} one can truncate small coefficients and reduce $d$.  Each data item to be loaded consists of the value of the coefficient $V_{pqrs}$ as well as the corresponding indices $p,q,r,s,\sigma,\tau$ that allow to apply the correct unitary $Q_{pq\sigma}Q_{rs\tau}$. QROAM allows to load these as qubit bitstrings. However, PREPARE requires the coefficient values as amplitudes of the state, not bitstrings. This gap is bridged by instead loading so-called ``keep-probabilities" and performing coherent alias sampling \cite{babbush_encoding_2018}. Fig.~\ref{fig:prepare original} shows a sketch of the PREPARE operator.

The number of data items $d$ to load can be reduced further by leveraging symmetries of the Hamiltonian that cause multiple identical coefficients. First of all, coefficient values are independent of spin. Thus we must only load one coefficient, and the other identical terms can be restored in the quantum circuit. 
Likewise, for a given permutation of orbital indices, eight coefficients which posses 8-fold symmetries (see Eqs~\eqref{eq: g_8fold}) can also be restored in the quantum circuit using only one set of orbital indices. This reduces the number of terms to be loaded $d$ by approximately a factor of 8.

The PREPARE operator is implemented in the following steps illustrated in Fig.~\ref{fig:prepare original} (see \cite{berry_qubitization_2019,lee_even_2021} for details):
\paragraph{Equal superposition state} Prepare $
        \frac{1}{\sqrt{d}} \sum_{l=0}^{d-1} \ket{l}
    $
    , where $d$ is the number of non-zero LCU terms (up to 8-fold and spin symmetries). This uses ancillas for amplitude amplification not shown in the figure.
\paragraph{Data loading} A QROAM loads data of width $m$ qubits. In principle, these $m$ qubits include the value of the coefficient indices $p(l),q(l),r(l),s(l)$ along with the value $w_l = |V_{p(l)q(l)r(l)s(l)}|/8$ and its sign $\theta$.
    
    However, in practice, in order to perform coherent alias sampling, slightly different data items must be loaded \cite{babbush_encoding_2018}.
    Instead of the coefficient value $V_{pqrs}$, a data field of $\aleph$ qubits (so-called ``keep-probability") is needed. $\aleph$ determines the accuracy with which  the coefficients $V_{pqrs}$ are ultimately loaded and can be computed with \eqref{eq:aleph}. Further, coherent alias sampling requires two values of the other data to be loaded (indices $p,q,r,s$ and $\theta$ and a qubit not mentioned here to distinguish between one- and two-body terms) \cite{babbush_encoding_2018}.
    Thus,
    \begin{equation}
      m = \aleph + 2(4\lceil\log P \rceil +2)
    \end{equation}
    where $P$ is the number of spatial orbitals ($2P$ is the number of spin orbitals).
    
    The QROAM is the asymptotically most expensive step with Toffoli cost
    \begin{equation}
        \lceil d/\kappa\rceil + m(\kappa-1) \label{eq:qroam cost}.
    \end{equation}
     Adjusting $\kappa$ (which must be a power of 2) leads to a tradeoff between Toffoli cost and ancilla qubit count \cite{Low2018Trading}
    \begin{equation}
        m\kappa + \lceil \log (d/\kappa) \rceil.
        \label{eq:qroam ancilla cost}
    \end{equation}
    Choosing $\kappa\sim\sqrt{d}$ to optimise Toffoli cost, both Toffoli and ancilla count of the data lookup asymptotically follow (dropping logarithmic factors)
    \begin{equation}
        \sim O(\sqrt{d})\sim O(P^2).
    \end{equation}
    While the QROAM lookup will also have to be uncomputed in UNPREPARE, the cost is lower because it doesn't depend on the size $m$ of the data items when using a measurement-based uncomputation scheme \cite{berry_qubitization_2019}.
 \paragraph{Coherent alias sampling} From the information thus loaded, coherent alias sampling then creates the state
    \begin{equation}
        \sum_{l=0}^{d-1} \sqrt{\frac{w_i}{\lambda}} \ket{l}\ket{0/1} \ket{p}\ket{q}\ket{r}\ket{s}\ket{\theta}\ket{\textrm{junk}_l}, \label{eq:subprepare}
    \end{equation}
    which is entangled to some $\ket{\textrm{junk}_l}$ that is not relevant. The second register, 0 or 1, flags one- or two-body terms, respectively.
 \paragraph{Symmetry restoration} Now, the spin symmetry and 8-fold symmetry must be restored. Two new qubits encoding spin $\ket{\sigma}$ and $\ket{\tau}$ in the state $(\ket{0} + \ket{1})/\sqrt{2}$ are added as further tensor product factors. When the tensor product is expanded, it quadruples the number of states in the superposition \eqref{eq:subprepare}.
    
To restore 8-fold symmetry, similarly three qubits $(\ket{0} + \ket{1})/\sqrt{2}$ for each of the symmetries $p\leftrightarrow q, r\leftrightarrow s, (p,q)\leftrightarrow(r,s)$ are added as tensor product factors. Swaps controlled on these qubits then swap $p,q,r,s$ registers depending on the symmetry.

The result is a state describing the full LCU in \eqref{eq:sparse LCU}.
A subtlety of symmetry restorations is that slightly different values of $w_i$ must be loaded by the QROAM, because the symmetry restoration accumulates factors of $1/\sqrt{2}$. Yet this does not affect the overall subnormalisation $\lambda$ of the Hamiltonian.

Next, a SELECT operator selects the correct unitary for the $p,q,r,s,\sigma$ and $\tau$ indices. The UNPREPARE operator uncomputes PREPARE. Using measurement based uncomputation, this is much more efficient than PREPARE \cite{berry_qubitization_2019}.
The total leading order Toffoli cost is 
\begin{equation}
    \left\lceil \frac{\pi\lambda}{2\epsilon_{\rm QPE}} \right\rceil \cdot \left(\left\lceil \frac{d}{\kappa}\right\rceil + m(\kappa-1)\right),
    \label{eq:dominant cost}
\end{equation}
the product of the QROAM cost for data loading \eqref{eq:qroam cost} with the number of iterations of the walk operator $\lceil \pi\lambda/(2\epsilon_{\rm QPE})\rceil$. The normalisation factor, $\lambda$, together with the desired accuracy, $\epsilon_{\rm QPE}$, determine the number of iterations of the walk operator required for phase estimation.

\subsubsection{Generalization of Sparse Qubitization for Bloch Basis Functions}~\label{sec:Bloch sparse qubitisation}

While the original sparse qubitization method supports Hamiltonians with real electron repulsion integrals only, Bloch orbitals usually lead to complex coefficients.
We generalise the sparse method to complex Hamiltonians by expanding the Hamiltonian in Majorana strings~\ref{sec:majorana representation} and instead of working with 8-fold symmetry restoration circuits, we introduce Majorana type restoration circuits. For a real Hamiltonian the expansion~\eqref{eq:1body re}--\eqref{eq: C-C-C} only contains the terms \eqref{eq:1body re}, \eqref{eq:2body B}, and \eqref{eq: 2body BBF}. The other terms arise for complex Hamiltonians. The coefficients of the Majorana strings are all real (or purely imaginary for the one-body terms) due to the Hamiltonian's Hermeticity.

\begin{figure}[H]
\centering
\includegraphics[width=0.95\textwidth]{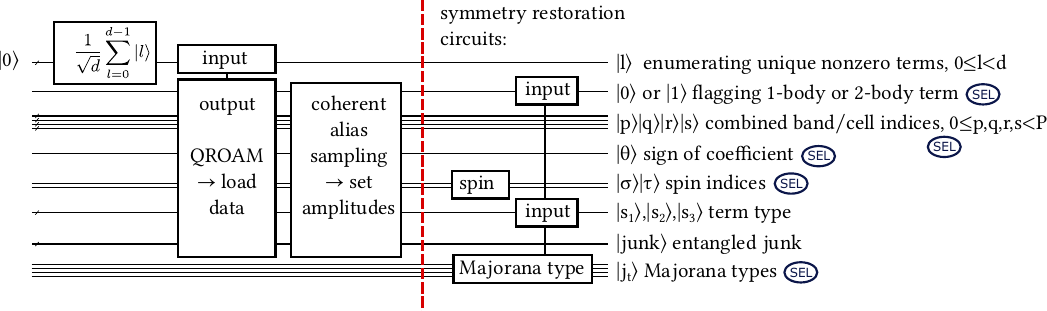}
\caption{PREPARE operator for the Bloch basis algorithm. In addition to the PREPARE operator of the original sparse qubitization (see Fig~\ref{fig:prepare original}), it has Majorana type indices that are used by SELECT to apply the correct Majorana types for the Majorana string described by a product state. $j_t=0,1$ and $t=1,2,3,4$ enumerates the positions of the Majorana operators in a Majorana string. The circuit for Majorana type symmetry restoration is given in the circuit~\eqref{eq:maj type restore circuit}; its input qubits must be loaded by QROAM. The spin symmetry restoration circuit is given in circuit~\eqref{eq:spin restore circuit}. The Hamiltonian has been written such that no 4-fold or 8-fold symmetry is present any more (see Sec.~\ref{sec:majorana representation}). To simplify the sketch we have depicted that both spin qubits are loaded by QROAM (See the circuit~\eqref{eq:spin restore table} for details).}
\label{fig:prepare bloch}
\end{figure}

We use a SELECT operator \cite{babbush_encoding_2018,von_burg_quantum_2021} that allows to select Majorana strings based on: the indices $p,q,r,s$; spin indices $\sigma,\tau$; and four Majorana type indices $j_1$ through $j_4$---as they appear in the Hamiltonian (section~\ref{sec:majorana representation}). In addition, the correct sign of the LCU coefficient is selected based on a $\ket{\theta}$ qubit. In Fig.~\ref{fig:prepare bloch}, the control qubits for SELECT are indicated by \includegraphics[height=2ex]{Figure_0.pdf}. The indices $p,q,r,s$ index the Bloch basis functions \eqref{eq:Bloch functions}, and as such they are composite indices, each consisting of band index and $k$-wave-vector index. For
the Bloch basis we do not need to split up the composite indices and arbitrarily enumerate them as 
\begin{equation}
    p, q, r, s \in \{0, \dots, P - 1\}.
\end{equation}

A main benefit of using Bloch basis functions even in classical methods is that momentum conservation causes many
terms to be zero (see \eqref{eq:Bloch functions}). Therefore we can expect the number of non-zero terms to scale as
\begin{equation}
d \sim O(M^4N^3),
\label{eq:bloch d}
\end{equation}
for an LCAO basis set, while for PW basis sets $M^4$ can be reduced to $M^3$.
Since the number of bands $M$ is defined per unit cell, the scaling of the algorithm with the system size, $N$, is cubic. 

The PREPARE operator is sketched in Fig.~\ref{fig:prepare bloch}.
The electron repulsion integrals in the original sparse method possess 8-fold symmetry in the $p,q,r,s$ indices, and this is restored in PREPARE with controlled SWAPS (see Sec.\ref{sec:original sparse qubitisation}). In our case, at first, the electron repulsion integrals merely have 4-fold symmetry \eqref{eq: g_4fold} because the basis functions are complex. Once the Hamiltonian is expanded in Majorana strings (section~\ref{sec:majorana representation}), this results in different types of symmetry for different coefficients as explained in sec~\ref{sec:majorana representation}.
Instead of restoring these symmetries with controlled SWAPS, we rewrite the LCU decomposition of the Hamiltonian such that the symmetries are not explicitly present anymore, see Sec.~\ref{sec:majorana representation}. The sums can be restricted to one branch of the symmetry by instead summing over the Majorana type. For example, the coefficient $B_{pqrs}$ in \eqref{eq:2body B} has 8-fold symmetry. Yet the sum is restricted to $p\le q, r\le s, (p,q)\le (r,s)$, such that only one branch of the symmetry is present in the LCU, and it does not have repeated coefficients for 8-fold permutations of $p,q,r,s$. The ``missing" terms are compensated by summing over multiple values of the Majorana type indices $j_1,\dots,j_4$. The resulting symmetry in the Majorana type can then be more easily restored, because it does not involve CSWAPS of multi-qubit $p,q,r,s$ registers. Further, even if we did not restrict the sums in this fashion, some Majorana type symmetry would be still present and have to be restored anyway due to the complex nature of the Hamiltonian.

The LCU has multiple repeated coefficients for different values of $j_1,\dots,j_4$ of the Majorana type indices, up to signs like $(-1)^{j_1+j_4}$. Rather than repeatedly loading the same coefficient with a different sign multiple times, we restore the Majorana type in PREPARE with the following symmetry restoration circuit:
 
\begin{equation}
    \begin{tikzpicture}
    \begin{yquant}
    qubit {1-body/2-body} a;
    qubit {$\ket{s_1}$} s1;
    qubit {$\ket{s_2}$} s2;
    qubit {$\ket{s_3}$} s3;
    qubit {$\ket{j_1}$} j1;
    qubit {$\ket{j_2}$} j2;
    qubit {$\ket{j_3}$} j3;
    qubit {$\ket{j_4}$} j4;

    h j1;

    x j2 ~ a, s1;
    cnot j2 | j1 ~ a;
    z j1 ~ a, s1;
    
    x j2, j4 | a ~ s1, s2, s3;
    cnot j2 | j1, a ~ s1, s2, s3;
    h j3 | a ~ s1, s2, s3;
    cnot j4 | j3, a ~ s1, s2, s3;
    z j1,j4 | a ~ s1, s2, s3;

    cnot j2 | j1, a, s3 ~s1, s2;
    h j3 | a, s3 ~ s1, s2;
    cnot j4 | j3, a, s3 ~ s1, s2;
    z j1,j4 | a, s3 ~ s1, s2, s3;

    x j2 | a, s2 ~ s3, s1;
    cnot j2 | j1, a, s2 ~ s3, s1;
    h j3 | a, s2 ~ s3, s1;
    cnot j4 | j3, a, s2 ~s3, s1;
    z j1 | a, s2 ~ s3, s1;

    x j3, j4 | a, s2, s3 ~ s1;
    cnot j2, j3, j4 | j1, a, s2, s3 ~ s1 ;

    cnot j2, j3, j4 | j1, a, s1 ~ s2, s3;

    cnot j2, j3, j4 | j1, a, s1, s3 ~ s2;
    x j1 | a, s1, s3 ~ s2;
    z j1 | a, s1, s3 ~ s2;
    \end{yquant}
    \end{tikzpicture}\label{eq:maj type restore circuit}
\end{equation}
The circuit needs three input qubits $\ket{s_1}\ket{s_2}\ket{s_3}$ along with the qubit flagging one-body or two-body terms that together distinguish all of the types of terms and thereby Majorana type symmetry in the LCU. The four output qubits $\ket{j_i}$ are initialised as $\ket{0}$ and will indicate Majorana type to be used by SELECT (see Fig.~\ref{fig:prepare bloch}). The initial values of the $\ket{s_i}$ qubits must be loaded from QROAM\footnote{The symmetry restoration circuit will add some factors of $1/\sqrt{2}$, similarly to spin restoration. In the Hamiltonian LCU (section~\ref{sec:majorana representation}), the coefficients in between the two $\sum$s are the values that must be loaded and prepared by coherent alias sampling. The factor in front of the first $\sum$ and the sign at the back of the terms are recovered along with the various symmetry restoration circuits.}.
They identify terms in the Hamiltonian as follows: 

\begin{equation}
    \begin{array}{rccc}
    \text{term} & \ket{s_1} & \ket{s_2} & \ket{s_3} \\\hline
    \text{1-body Re} &  \ket{0} & \ket{0} & \ket{0}\\
    \text{1-body Im} &  \ket{1} & \ket{0} & \ket{0} \\
    B & \ket{0} & \ket{0} & \ket{0}  \\
    F &  \ket{0} & \ket{0} & \ket{1} \\
    C & \ket{0} & \ket{1} & \ket{0} \\
    F-B+B & \ket{0} & \ket{1} & \ket{1} \\
    F-F+F & \ket{1} & \ket{0} & \ket{0} \\
    C-C+C & \ket{1} & \ket{0} & \ket{1} \\
    \end{array}
    \label{eq:maj type restore table}
\end{equation}

Let us give an example of Majorana type symmetry restoration for a term of type $B$, i.e.~\eqref{eq:2body B}. The table above shows that the qubits are loaded such that at the vertical dashed line (red) in Fig.~\ref{fig:prepare bloch} the input qubits are
\begin{equation}
 \ket{\text{1 or 2 body}}\ket{s_1}\ket{s_2}\ket{s_3} = \ket{1}\ket{0}\ket{0}\ket{0}.
\end{equation}
The symmetry restoration circuit \eqref{eq:maj type restore circuit} acting on these then results in the state
\begin{equation}
\ket{1000} \otimes \frac{1}{2}\left(-\ket{0101} +\ket{1001} + \ket{0110} -\ket{1010}\right) = \ket{1000} \otimes \frac{1}{2} \sum_{j_1\neq j_2, j_3\neq j_4} \ket{j_1j_2j_3j_4} (-1)^{j_1+j_4}.
\label{eq:maj restoration example}
\end{equation}
This selects the correct Majorana type indices and signs for the $B$-type term in \eqref{eq:2body B}.
Note that when inverting the circuit for UNPREPARE, one must omit the CZs. Otherwise the $-1$ in the Majorana symmetries of the coefficients (like in \eqref{eq:maj restoration example}) introduced in both PREPARE and UNPREPARE would cancel. The cost of this circuit is subleading, except to the extent that the necessary data loading increases $m$. Its cost can be reduced by using a unary iteration over the $\ket{s_i}$ qubits.

As in the original sparse method, the spin symmetry is also restored on the quantum computer. Identical coefficients corresponding to different spin configurations must only be loaded once. Here we have two spin qubits $\ket{\sigma}$ and $\ket{\tau}$ for the first two and second two Majoranas, respectively. Because we have more possible combinations than in the original sparse method, we need a short spin symmetry restoration circuit
\begin{equation}
\begin{tikzpicture}
\begin{yquant}
qubit {$\ket{\sigma}$} sigma;
qubit {$\ket{\tau}$} tau;
h sigma;
cnot tau | sigma;
\end{yquant}
\end{tikzpicture}
\label{eq:spin restore circuit}
\end{equation}
Similarly to Majorana restoration, $\ket{\sigma}$ is initialised as $\ket{0}$ and different initial values for $\ket{\tau}$ must be loaded for different coefficient types:
\begin{align} \label{eq:spin restore table}
\ket{0} &\ \text{for 1-body terms and 2-body terms with}\ \sum_\sigma\ \text{and} \\
\ket{1} &\ \text{for 2-body terms with}\ \sum_{\sigma\neq \tau}. \nonumber
\end{align}
Note that (contrary to the simplified Fig.~\ref{fig:prepare bloch}), the qubit $\ket{\sigma}$ is not loaded from QROAM but initialised as zero.

As in the original sparse qubitization algorithm, the dominant cost is \eqref{eq:dominant cost}, the product
of the number of iterations in phase estimation and the QROAM cost of data loading. Here, the parameters are as follows:
\begin{itemize}
    \item $\epsilon_{\rm QPE}$ is the error budget for the phase estimation.
    \item $\lambda$ is the normalization factor of the Hamiltonian, i.e.~the $L_1$ norm of the LCU (section~\ref{sec:majorana representation}). It depends on the specific material under consideration.
    \item $d$ is the number of non-zero coefficients of the LCU to load from QROAM. (Up to spin symmetry and Majorana type symmetry, which are restored with circuits as discussed above.)
    \item $m$ is the size of each of the $d$ data items that need to be loaded. Specifically, we have
    \begin{equation}
        m = \aleph + 2 \left(  4\lceil\log (P)\rceil + 6  \right)
    \end{equation}
    The values of the coefficients $V_{pqrs}$ are effectively encoded in the $\aleph$ qubits and restored
    with coherent alias sampling \cite{babbush_encoding_2018}. The other qubits correspond to the necessary indices and further
    qubits to be loaded. For technical reasons of coherent alias sampling, they must be loaded with two values each, explaining the factor of 2 in above equation. First, $4\log(P)$ is the total width of the four basis function indices $p,q,r,s$. A small number of qubits are needed for Majorana
    type restoration, spin symmetry restoration, distinguishing one- and two-body terms, and the sign
    of the coefficient.
    \item $\kappa$ is a power of 2, and can be tuned to achieve a tradeoff between Toffoli and ancilla count \eqref{eq:qroam ancilla cost}. We choose it such that Toffoli count is minimized. Then the overall cost (dropping logarithmic factors) is
    \begin{equation}
        \sim O(\lambda\sqrt{d}/\epsilon_{\rm QPE}) \sim O(\lambda M^2N^{3/2}/\epsilon_{\rm QPE}).
        \label{eq:dominant cost scaling}
    \end{equation}
    
\end{itemize}

\subsubsection{Generalization of Sparse Qubitization for Wannier Basis Functions}~\label{sec:Wannier sparse qubitisation}

\begin{figure}[H]
\centering
\includegraphics[width=0.94\textwidth]{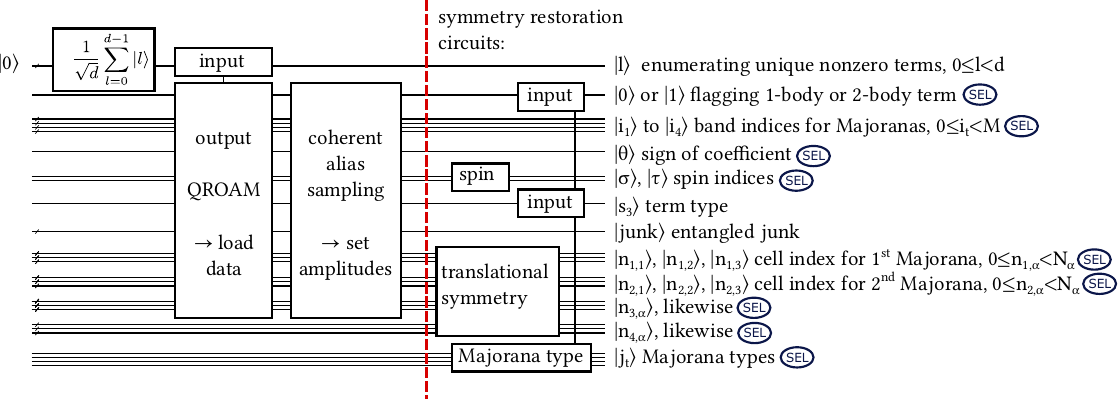}
\caption{PREPARE operator for the Wannier basis algorithm. We make use of translational symmetry of the LCU's coefficients \eqref{eq:Wannier integral to load} to reduce data loading. This necessitates splitting the combined indices $p,q,r,s$ into a band index $i_t$ and a cell index ${\bf n_t} = (n_{t,1},  n_{t,2}, n_{t,3})$. $t=1,2,3,4$ enumerates the positions of the Majorana operators in a Majorana string. $j_t=0,1$. The translational symmetry restoration circuit is given in the circuit~\eqref{eq:wannier in-place additions}.}
\label{fig:prepare wannier}
\end{figure}

The algorithm closely follows the one for Bloch states. However, since Wannier functions can be chosen to be real, only the terms \eqref{eq:1body re}, \eqref{eq:2body B} and \eqref{eq: 2body BBF} in LCU expansion from Sec.~\ref{sec:majorana representation} are non-zero. Additionally, these terms possess the same translational symmetry as ERIs~\eqref{eq: translational_symmetry_2}.
This is taken into account through a translational symmetry restoration circuit which further reduces the cost of quantum computation.
We sketch the PREPARE operator in Fig.~\ref{fig:prepare wannier}.

Translational symmetry can be leveraged to avoid loading repeated coefficient values. In order to restore it with a symmetry restoration circuit, the compound indices $p,q,r,s$ must be split into an orbital index $i_t$ and cell index $\tn_t = (n_{t,1}, n_{t,2}, n_{t,3})^T$ as indicated in Fig.~\ref{fig:prepare wannier}, $t=1,2,3,4$ enumerates the positions of the Majorana operators in a Majorana string. The figure also shows that SELECT must now be controlled on all qubits constituting the compound indices. While $i_t\in\{0,\dots,M-1\}$ for the number of orbitals per cell $M$, the cell index $n_{t,i}\in\{0,\dots,N_{i}-1\}$ for each spatial direction.
Even though the total number of non-zero terms is larger than in the Bloch basis set, the translational symmetry reduces the number of unique non-zero terms $d$ to the same asymptotic scaling
\begin{equation}
   d \sim O(M^4N^3)
\end{equation}
as in the Bloch basis \eqref{eq:bloch d}.

The spin restoration circuits work identically to circuits used for the Bloch representation, while the Majorana type symmetry restoration circuit \eqref{eq:maj type restore circuit} can be simplified:
Since the only terms appearing are the one-body term Eq.~\eqref{eq:1body re}, and two two-body terms~Eq.~\eqref{eq:2body B} and Eq.~\eqref{eq: 2body BBF}, we can remove all of the gates from \eqref{eq:maj type restore circuit} that are controlled for other contribution in the Hamiltonian. Further, we can remove the qubits $\ket{s_1}\ket{s_2}$ keeping only $\ket{s_3}$.

The translational symmetry of the LCU coefficients $B$ from Eq.~\eqref{eq:2body B} can be expressed as 
\begin{align}
    B_{i_1 \tn_1, i_2 \tn_2, i_3\tn_3,i_4\tn_4} &= B_{i_1 (\tn_1+{\bf v}), i_2 (\tn_2+{\bf v}), i_3 (\tn_3+{\bf v}),i_4(\tn_4+{\bf v})}
    = B_{i_1 (\tn_1-\tn_4), i_2 (\tn_2-\tn_4), i_3 (\tn_3-\tn_4),i_4{\bf 0}}, \quad
    \forall {\bf v}\in\vec{\mathbb{Z}} 
    \label{eq:Wannier integral to load}
\end{align}
with additions taken mod $(N_1, N_2, N_3)^T$ independently for each spatial component of $\tn$. (The same holds for one-body and the other two-body coefficients.) We reduce  the data loading by only loading one of these coefficients, and restoring the others with the following translational symmetry restoration circuit\footnote{In order to cancel the $1/\sqrt{N}$ coming from the Hadamards, the coefficients with $\tn_4=0$ should that are loaded should be multiplied $N$ times their value. The total subnormalisation of the Hamiltonian does not change.}: 
\begin{equation}
    \begin{tikzpicture}
    \begin{yquant}
    qubit {$\ket{\tn_1}$} k1;
    qubit {$\ket{\tn_2}$} k2;
    qubit {$\ket{\tn_3}$} k3;
    qubit {$\ket{\tn_4}=\ket{0}\otimes\ket{0}\otimes\ket{0}$} k4;
    
    slash k1, k2, k3, k4;
    
    H k4;
    [name=a1]
    box {\Ifnum\idx=1 In\Else add\Fi} k4, k1;
    [name=a2]
    box {\Ifnum\idx=1 In\Else add\Fi} k4, k2;
    [name=a3]
    box {\Ifnum\idx=1 In\Else add\Fi} k4, k3;

    \draw (a1-0) -- (a1-1) (a2-0) -- (a2-1) (a3-0) -- (a3-1);
    
    \end{yquant}
    \end{tikzpicture}
    \label{eq:wannier in-place additions}
\end{equation}
The in-place additions \cite{Gidney2018Halving} are to be performed separately for the three spatial directions of $\tn_t$. In this circuit, $\tn_4$ is an ancilla and the index must not be loaded from QROAM, leading to a reduction in $m$ compared to the Bloch basis. 

The dominant cost of the algorithm in the Wannier basis has the same formula 
\eqref{eq:dominant cost} and scaling \eqref{eq:dominant cost scaling} as in the Bloch basis. Yet, $m$ is lower in the Wannier basis. This reduces the dominant Toffoli \eqref{eq:qroam cost} and qubit costs \eqref{eq:qroam ancilla cost} which stem from data loading. Further, $d$ is expected to be lower (if only by a factor) because of the two-body terms, only Eq.~\eqref{eq:2body B} and Eq.~\eqref{eq: 2body BBF} are non-zero for real orbitals, and, as we will find in Sec.~\ref{sec:errors and truncation}, more small coefficients can be truncated. The dependence of the normalization factor, $\lambda$, which has a large effect on the dominant cost, on the choice of the basis set is discussed in Sec.~\ref{sec:results}.

\subsubsection{Error Correction Scheme}\label{sec:error_correction}

The details of the error correction implementation we used in this work can be found in Sec.~4 of Ref.~\cite{blunt2022perspective}, but for completeness a short description is also provided here. 

We estimate the error correction overheads following Litinski's Game of Surface Codes scheme~\cite{litinski_game_2019}. The computational qubits are arranged in \emph{fast blocks}~\cite{litinski_game_2019}, that enable the consumption of one magic state per time step -- the clock rate of an error corrected quantum processor. We select magic state factories from a subsect of the ones described in \cite{litinski_game_2019}\footnote{Specifically, we consider 15-to-1, 14-to-2, 20-to-4, 116-to-12, and 225-to-1 factories. The largest calculations all require the highest fidelity factory in this list: 225-to-1. \label{fn:factories_list}}. We allow ourselves as many magic state factories as necessary, so that computational qubits do not idle between magic state injections. Magic state factories are arranged around the computational qubits, accounting for the routing space necessary to move the distilled T states to the computational area. 

The code distance $d$ and magic state factory types are determined by allocating a failure rate budget of 1$\%$ to error correction; this budget is divided between magic state distillation failure ($0.1\%$) and probability of logical circuit failure ($0.9\%$).  Logical circuit failure is estimated using the conventional formula for error rate in the surface code \cite{fowler_low_2018}:
\begin{equation}
    P_{\textrm{logical}}=0.1\times(100\,p)^{\frac{d+1}{2}}\,,
\end{equation}
where $p$ is the probability of physical error, that we take to be $10^{-3}$ or $10^{-4}$.  The former is an accepted rate for superconducting systems, while the latter is considered optimistic in these platforms.

To obtain runtime from T-gate count, we assume serial implementation and multiply the number of T gates by the duration of the time step, which in the surface code is $d$ times the duration of the code cycle.  We take the code cycle (the time needed to carry out one round of stabiliser measurements) to be 1$\mu\textrm{s}$ long --
this is a conventional figure for superconducting qubits~\cite{krinner_realizing_2022,acharya_suppressing_2022}.

\subsection{Classical Computational Details}\label{subsec: classical_computational_details}
In order to generate ERIs, we have carried out restricted Hartree-Fock calculations in the Gaussian basis set using the PySCF software~\cite{PySCF1,PySCF2,PySCF3,PySCF4}. The Fermi-Dirac distribution of occupation numbers has been used. In order to numerically study the scaling properties of the algorithms, we carry out calculations on a model hydrogen (H) crystal in body-centre cubic (BCC) structure. Quantum resource estimations have also been applied to realistic solids such as lithium hydride (LiH), NiO and PdO.
In the case of H and LiH, all-electron calculations have been carried out with STO-3G~\cite{hehre1969a,hehre1970a,feller1996a,schuchardt2007a,pritchard2019a} basis set, while for NiO and PdO the GTH pseudopotentials~\cite{goedecker_separable_1996,hartwigsen_relativistic_1998} with GTH-SZV basis set have been used~\cite{vandevondele_quickstep_2005,vandevondele_gaussian_2007}. Gaussian density fitting has been employed for the calculation of ERIs~\cite{Sun_2017_gaussian_and_pw_df}. The pseudopotential setups with 18 valence electrons for Ni and Pd, and 6 valence electrons for O have been used.
The geometry of H has been optimized with GPAW software~\cite{Mortensen2005,Enkovaara2010,ASE}, the geometry of PdO has been optimized with Quantum Espresso~\cite{Espresso1,Espresso2}, while experimental structures have been used for LiH~\cite[p.7]{Messer_1960} and NiO~\cite{Sasaki_1979}. We provide geometry files in the Supplemental Information. In order to obtain spatially localized orbitals satisfying translational symmetry~\eqref{eq: Wannier symmetry}, the natural atomic orbitals~\cite{reed_natural_1985} in a supercell have been constructed. We have then validated that the translational symmetry~\eqref{eq: translational_symmetry_1},~\eqref{eq: translational_symmetry_2} of the integrals is satisfied (see Table~\ref{tab:translation_symmetry_error}). 
Calculations in the supercell have been carried out using cubic cells of different sizes for H, LiH and NiO, while a tetragonal cell has been used for PdO. For calculations using Bloch functions, we used primitive 2-atom cells for LiH and NiO with varying $k$-point meshes (up to a 3x3x3 mesh). The unit cell used for H consists of two atoms while the tetragonal unit cell of PdO has been chosen such that it contains four atoms.

\begin{table}[H]
    \centering
    \begin{tabular}{c|c}
        System &  Error (Ha)\\
        \hline
        H (16 atoms)  & $1.53 \times 10^{-11}$  \\ 
        H (54 atoms)  & $1.70 \times 10^{-6}$   \\ 
        H (128 atoms) & $3.53 \times 10^{-10}$ \\ 
        H (250 atoms) & $2.43 \times 10^{-10}$ \\ 
        LiH (64 atoms) & $3.08 \times 10^{-9}$ \\ 
        NiO (64 atoms) & $9.76 \times 10^{-8}$ \\ 
        PdO (16 atoms) & $5.11 \times 10^{-11}$ \\
        PdO (72 atoms) & $2.41 \times 10^{-5}$ \\ 
    \end{tabular}
    \caption{The magnitude of the translational symmetry error, $\epsilon_t$, in one body terms arising from generation of localized orbitals in the supercell calculations. $\epsilon_t = \max_{ij \tR \tR'} | h_{\tR i, \tR' j} - h_{0i, \tR'-\tR j} |$.}
    \label{tab:translation_symmetry_error}
\end{table}

\subsubsection{Errors and Truncation Strategies}
\label{sec:errors and truncation}

The total additive error of the energy estimation can be split into three errors~\cite{lee_even_2021}:
\begin{equation}\label{eq: total_error}
    \epsilon = \epsilon_{\rm QPE} + \epsilon_{\rm trunc} + \epsilon_{\rm prep},
\end{equation}
where $\epsilon_{\rm QPE}$ is the error due to limited precision of the QPE,  $\epsilon_{\rm trunc}$ is the error due to discarding Hamiltonian coefficients with a certain precision, and $\epsilon_{\rm prep}$ is the error due to finite precision of state preparation amplitudes. Given the accuracy $\epsilon$ we distribute the total error as
\begin{align}
    \epsilon_{\rm QPE} = \frac{10}{16} \epsilon, \quad
    \epsilon_{\rm trunc} = \frac{3}{16} \epsilon, \quad
    \epsilon_{\rm prep} = \frac{3}{16} \epsilon.
\end{align}
Since in condensed matter calculations one is often interested in the energy of the crystal per formula unit (f.u.), we allow the total error, $\epsilon$, to grow proportionally to the system size, that is the error in energy estimation per f.u.~is fixed when the size of the supercell increases. The required total accuracy of calculations strongly depends on the problem of interest. For example, the band gap of transition metal oxides is on the order of a few eV, while the energy difference between different crystal structures can be up to a few hundred meV/f.u.~\cite{peng_2017_synergy_vdW}. Therefore, we carry out truncation with different accuracy of 50 meV/f.u.~(1.8 mHa/f.u., which is close to chemical accuracy), 5 meV/f.u., and 0.5 meV/f.u.

\begin{figure}[H]
\centering
\includegraphics[width=0.45\textwidth]{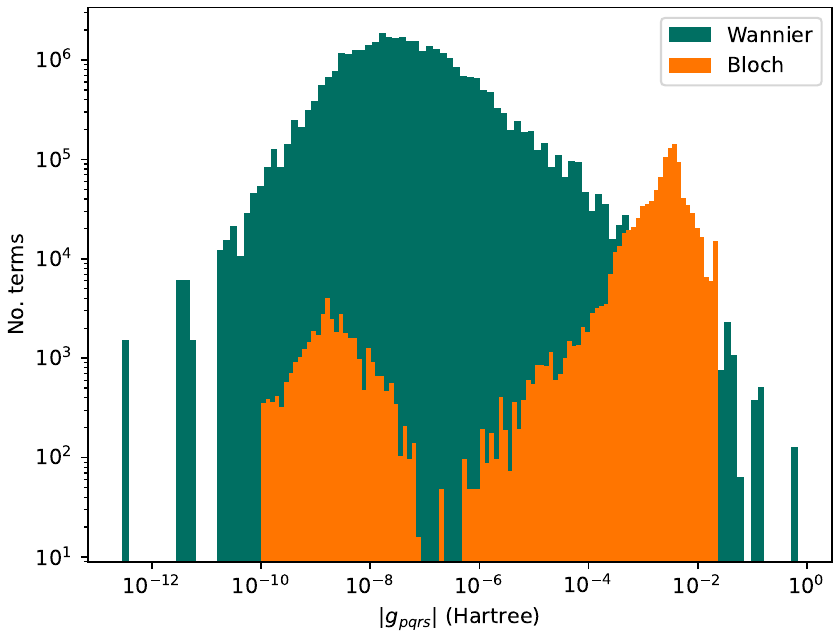}
\caption{\label{fig: BvsW-no-terms} The absolute value distribution of two-body matrix elements for H using Bloch and Wannier functions.
The Hamiltonian terms were truncated at $10^{-16}$ Ha. The $k$-point mesh of $(4, 4, 4)$ is used for Bloch functions with a 2-atom unit cell, and a 128-atom supercell is used for constructing Wannier functions.
}
\end{figure}

One of the bottlenecks in the quantum algorithm is the number of data items, $d$, that must be loaded into a quantum computer. The number of data items can be reduced by truncating small terms in the Hamiltonian. In order to estimate the error $\epsilon_{\rm trunc}$ due to truncation of the Hamiltonian coefficients, one usually exploits classical heuristics based on the norm of the Hamiltonian's coefficients~\cite{von_burg_quantum_2021} or classical quantum chemistry calculations such as coupled-cluster theory~\cite{berry_qubitization_2019,lee_even_2021}. For large systems, such as crystalline solids, high-order coupled-cluster calculations require large computational efforts and, therefore, we have used a simpler truncation methods based on the LCU's $L_2$ norm following Ref.~\cite{von_burg_quantum_2021}. Namely, if $H = \sum_i w_i U_i$ as in Eq.\eqref{eq:LCU} then we define the truncated Hamiltonian $H_{\rm trunc} = \sum_i w^{\rm trunc}_i U_i$ based on the following equation:
\begin{equation}\label{eq: L2-norm-trunc}
   \sqrt{\sum_i |w_i - w^{\rm trunc}_i|^2} < \epsilon_{\rm trunc}  
\end{equation}
One can also estimate the error based on $L_1$-norm truncation and even though such an estimation is rigorous, it does not allow truncating many coefficients. As shown in Fig.~\ref{fig: BvsW-no-terms}, there are many two-body terms in the Wannier representation with magnitude less then $10^{-8}$ Ha, and as a result, the error estimate based on $L_1$-norm would already be on the order of $0.01$ Ha if all these terms are neglected. The $L_2$-norm allows truncating much more coefficients but it is not a rigorous error estimate and, therefore, it might provide optimistic resource estimations. Instead of using $L_1$-norm-based truncation, we will also present results for a truncation threshold of $10^{-9}$ Ha in order to understand how much $L_2$-norm truncation reduces resource as compared to an accurate representation of the Hamiltonian. We use this procedure for moderate size systems. However, large systems such as NiO and PdO with supercells made of 64 and 72 atoms with around 900 spin orbitals require large amount  of memory. In this case, we save only those integrals whose absolute value is less then $10^{-7}$ Ha and then the Hamiltonian coefficients in Majorana representation are truncated with the threshold ($1.81\times10^{-6}$ Ha and $3.82\times10^{-7}$ Ha for NiO-64 and PdO-72, respectively) obtained from calculations on smaller supercells at a total accuracy, $\epsilon$, of 5 meV/f.u.

\section{Results}\label{sec:results}

\subsection{Scaling}

\begin{figure}[H]
\centering
\includegraphics[width=0.9\textwidth]{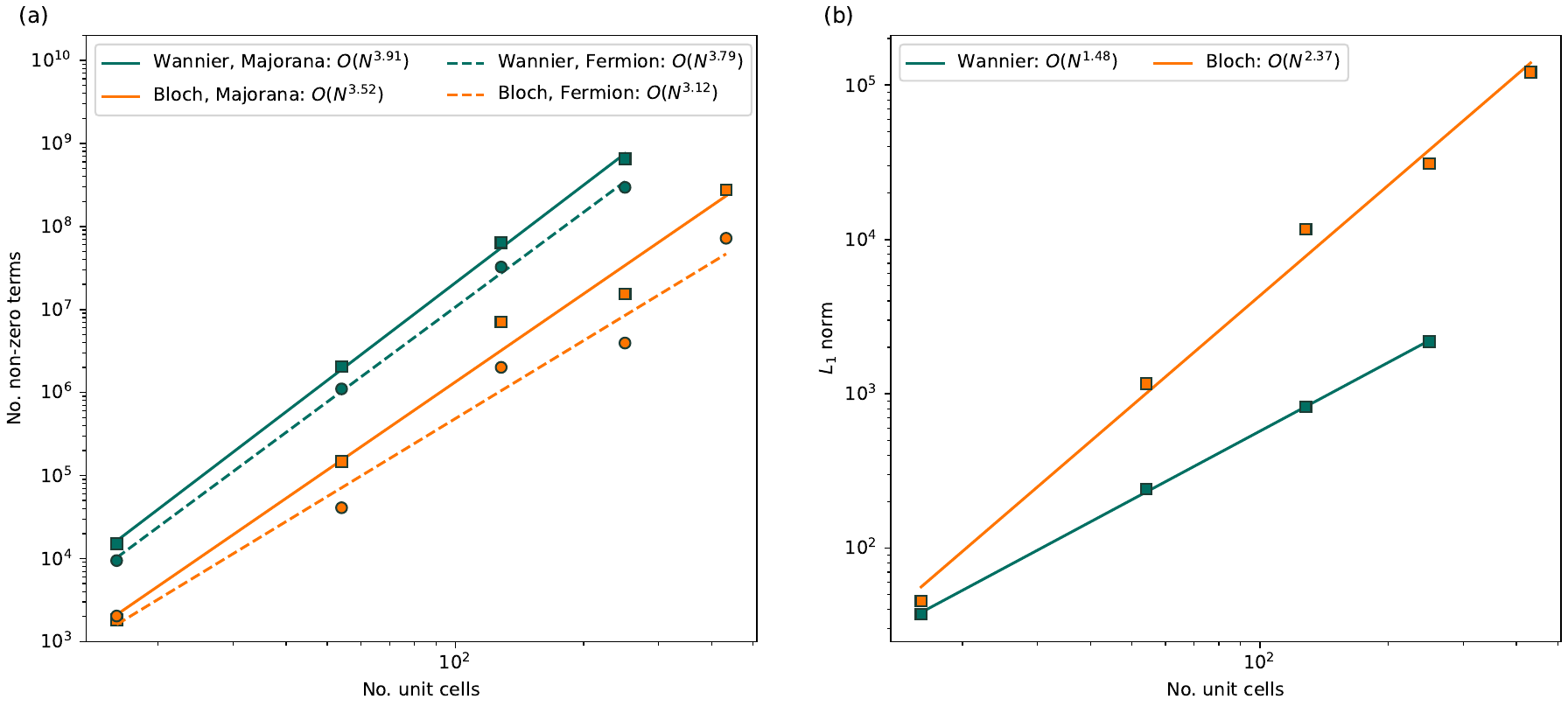}
\caption{\label{fig: 3d-Hydrogen-Ham} (a) The total number of non-zero terms before taking into account translational symmetry of Wannier functions or symmetry of Brillouin zone and (b) the $L_1$ norm of the Hamiltonian in Majorana representations. Hydrogen in BCC lattice structure with varying supercell size or $k$-point sampling ($x$-axis) is used.
The Hamiltonian terms were truncated at $10^{-9}$ Ha.
}
\end{figure}

\begin{figure}[H]
\centering
\includegraphics[width=0.9\textwidth]{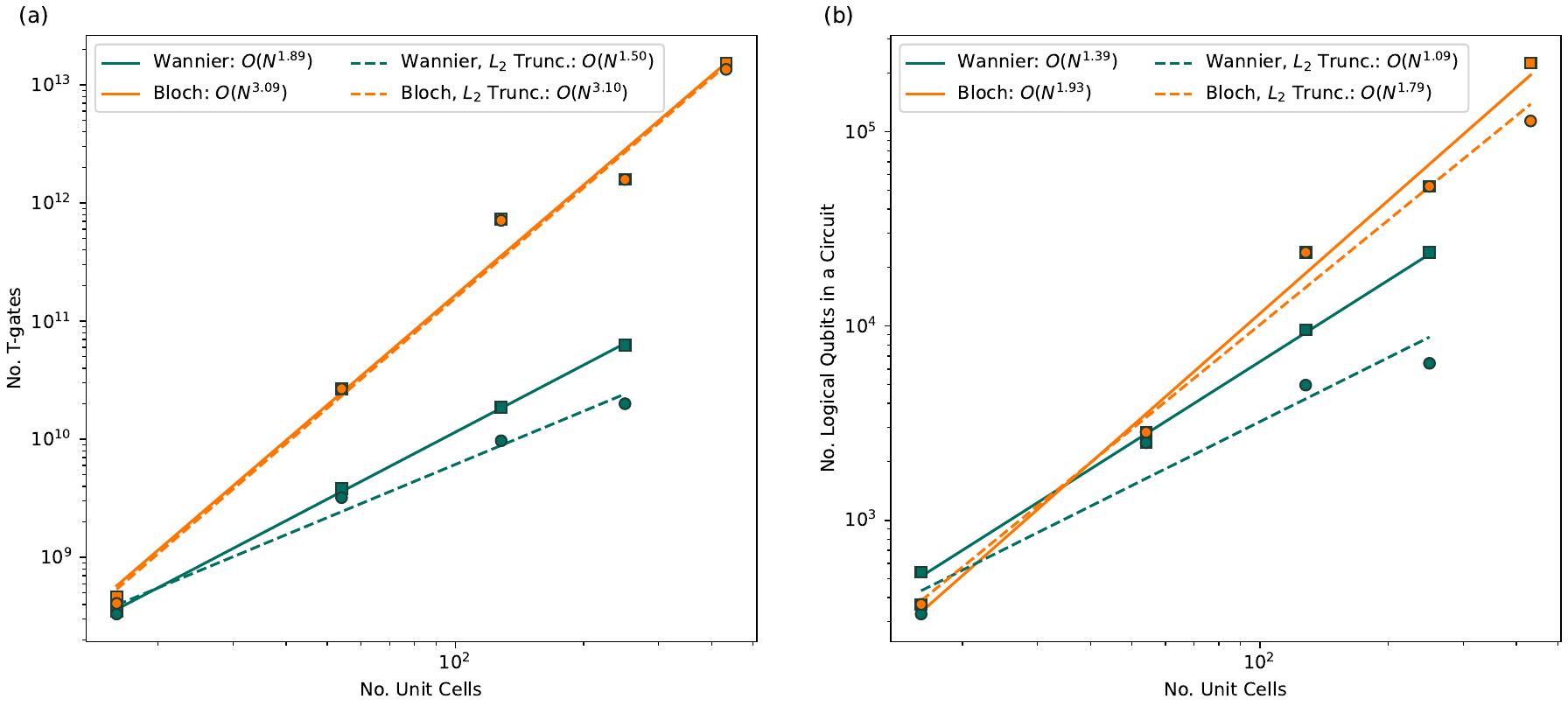}
\caption{\label{fig: 3d-Hydrogen-T-gates-Lq} (a) The number of T gates (b) The number of logical qubits in the circuit; no error correction has been applied here. Hydrogen in BCC lattice structure with varying supercell size or $k$-point sampling ($x$-axis) is used. Resource estimations carried out so that the total accuracy, $\epsilon$, is 0.5 meV/f.u.
}
\end{figure}

Both Wannier and Bloch functions provide the same spectrum of the Hamiltonian since they are related to each other through unitary transformation. However, the Hamiltonian have different properties in different basis sets and the main task is to chose such basis functions which minimize the total cost of quantum computation. It is proportional to the square root of the number of Hamiltonian coefficients, $d$,  loaded into the quantum computer times the $L_1$ norm, $\lambda$, of the Hamiltonian divided by the precision of QPE simulation $\epsilon_{\rm QPE}$ (see Eq.~\eqref{eq:dominant cost scaling}). With the example of model H in a BCC lattice structure, one can see that the number of non-zero terms in the Hamiltonian in Bloch representation scales as a cubic power of the system size, which is almost an order of magnitude better than in Wannier representation if no symmetries are taken into account (see~Fig.~\ref{fig: 3d-Hydrogen-Ham}(a)). However, the trend is opposite for the $L_1$ norm, $\lambda$, and the Hamiltonian has a much smaller norm using Wannier functions (see~Fig.~\ref{fig: 3d-Hydrogen-Ham}(b)) which in turn significantly reduces the number of controlled-unitary steps in Hamiltonian simulation. This is on par with molecular calculations where the set of localized orbitals reduces the $L_1$ norm~\cite{Koridon2021}. The number of T gates shown on Fig.~\ref{fig: 3d-Hydrogen-T-gates-Lq} is significantly lower in Wannier representation and has a better asymptotic scaling. This is achieved by (i) loading only unique Hamiltonian terms into the quantum computer as discussed in~\ref{sec:Wannier sparse qubitisation}, and (ii) providing a lower $L_1$ norm, $\lambda$, than in Bloch representation. Further reduction in T-count can be accomplished by truncating the Hamiltonian coefficients using the $L_2$-norm based truncation as shown on Fig.~\ref{fig: 3d-Hydrogen-T-gates-Lq}. After applying the $L_2$-norm truncation, the number of T gates in model H systems scales as $O(N^{1.5})$ with the system size. Such a low asymptotic scaling can also be explained by the fact that the permissible error grows proportionally to the system size and the number of terms, remained after truncation in such model systems with Wannier functions, scales as $O(N^3)$. In molecular systems, one can expect that localized orbitals produce Hamiltonians with quadratic scaling w.r.t. number of non-zero terms~\cite{sabzevari_improved_2018}. We can expect the same for materials with large band gaps while the model H system in our calculations have zero-gap at both PBE and Hartree-Fock level of theory.

\subsection{Resource Estimations}

\begin{figure}[H]
\centering
\includegraphics[width=1.0\textwidth]{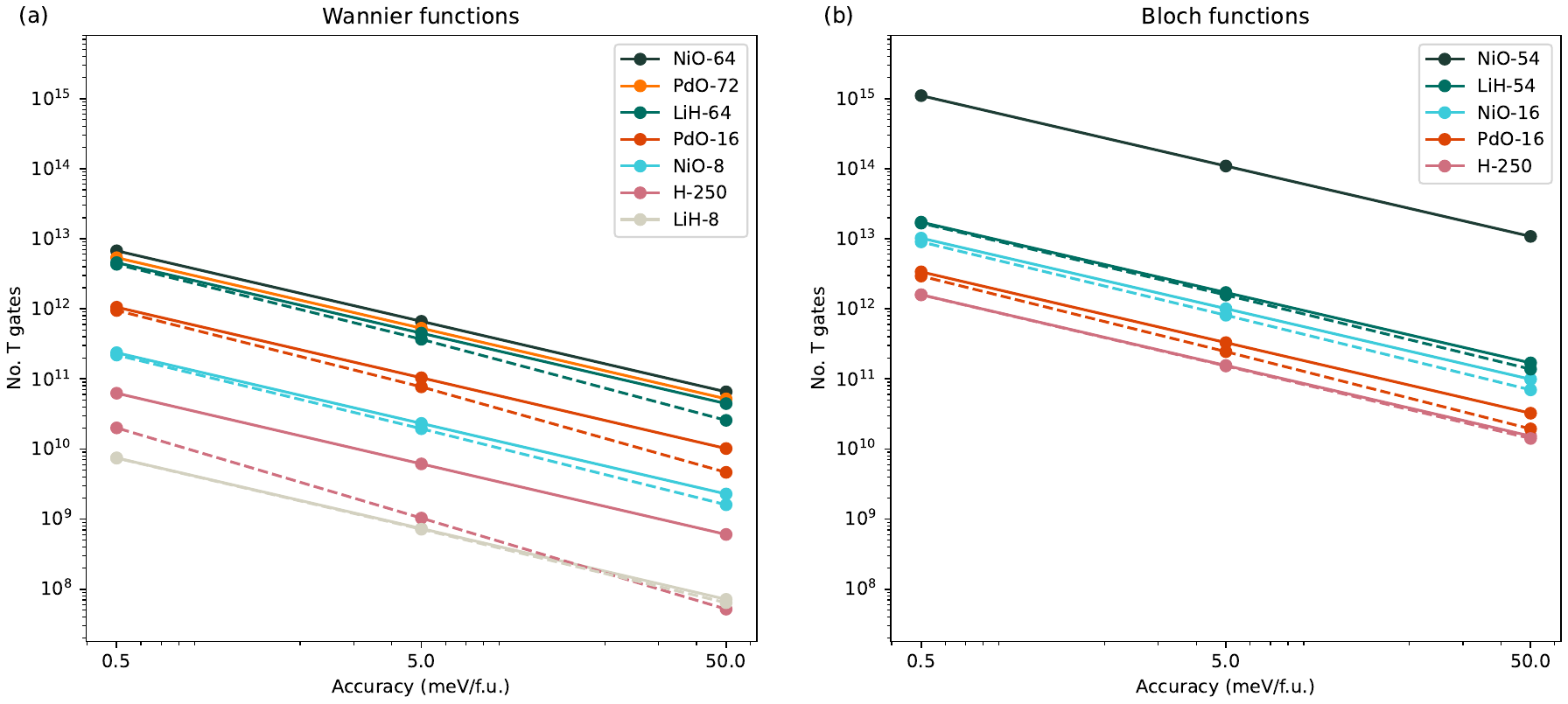}
\caption{\label{fig: T_count_WvsB_fixed_and_l2} 
The number of T gates as a function of the total accuracy of the Hamiltonian simulation, $\epsilon$, using different basis functions: (a) Wannier functions. (b) Bloch functions. Solid lines have been obtained for Hamiltonians in which coefficients are truncated at $10^{-9}$ Ha except for NiO-64 and PdO-72, where truncation was chosen based on smaller cell calculations due to high memory requirements. Dashed lines were obtained using adaptive truncation threshold based on $L_2$ norm as explained in the main text. Legends indicate the name of the materials and the number next to the name is the total number of atoms in the supercell or the size of the $k$-point mesh for Wannier and Bloch functions, respectively.
}
\end{figure}

The number of T gates required for a single shot of the qubitized QPE circuit is presented in Fig.~\ref{fig: T_count_WvsB_fixed_and_l2}. When Wannier functions are used as a basis set and the targeted accuracy is 50 meV/f.u., the number of T gates in the circuit is less than $10^{11}$ for all materials, system sizes and truncation strategies used in this work. As one can see, reducing the total permissible error of the Hamiltonian simulation by a factor of 10 increases the number of T gates by approximately an order of magnitude in agreement with Eq.~\eqref{eq:dominant cost scaling}. The L$_2$-norm truncation strategy reduces the number of T gates but not much except for H in Wannier basis, where the truncation of coefficients leads to an order of magnitude reduction of T gates. Similar trends are observed when Bloch functions are employed. However, in this basis set, the T-gate count is consistently higher than the T-gate count obtained with Wannier functions: for H the number of T gates is 25 times larger, for LiH with 54 $k$-points the T count is four times larger than for LiH with 64 atoms in the supercell whereas for NiO the difference between the two approaches is almost two orders of magnitudes. We can see that more than an order of magnitude comes from the fact that the L$_1$ norm in Wannier representation is smaller and the other improvement comes from truncation of small coefficients and symmetry considerations.

\begin{figure}[H]
\centering
\includegraphics[width=1.0\textwidth]{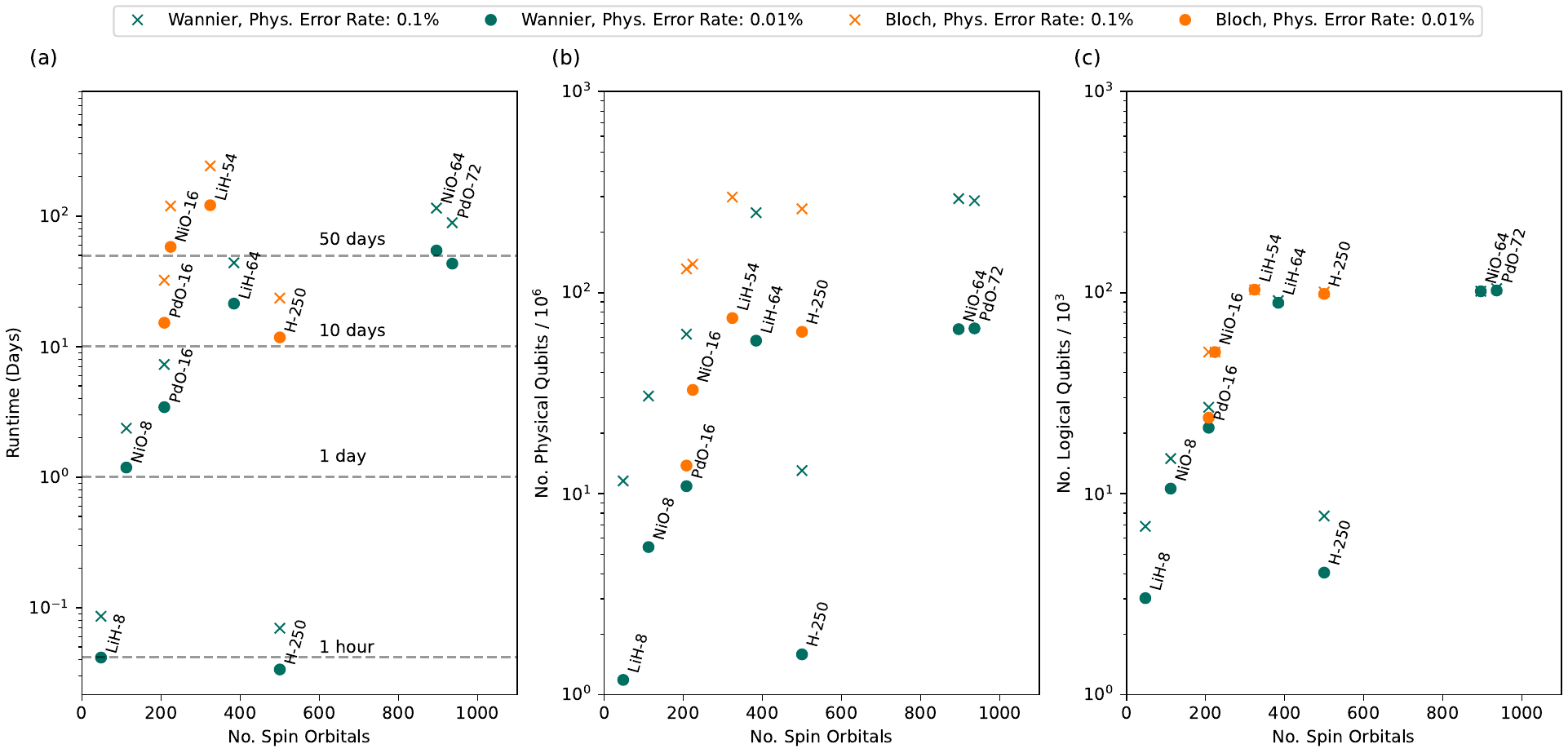}
\caption{\label{fig: Runtime_Logical_Phys} 
Resource estimation for calculation of the ground state energy of several solid state systems using Wannier and Bloch functions. The results have been obtained based on L$_2$-norm truncation and the accuracy of Hamiltonian simulations is 50 meV/f.u. (1.8 mHa/f.u.). 
The code cycle duration is $10^{-6}$ s. The physical error rates are assumed to be $0.1\%$ and $0.01\%$.
(a) Runtime in days (b) The number of physical qubits (c) The number of logical qubits. $x$-axis is the total number of spin orbitals. The number next to the name of the crystal indicates how many atoms in the supercell contains and how large the $k$-point mesh has been used for Wannier and Bloch basis sets, respectively.
} 
\end{figure}

Fig.~\ref{fig: Runtime_Logical_Phys}(a) demonstrates the runtime required for a single shot of the QPE circuit with permissible total error of 50 meV/f.u. As can be seen, small-unit-cell simulations of NiO and PdO consisting of 8 and 16 atoms, respectively, can be performed within less than 10 days. Systems with larger computational cells such as LiH with 64 atoms in the cell can be simulated within 50 days even if the physical error rate of gate operations is 0.1\%. Materials such as NiO with 64 atoms and PdO with 72 atoms in the supercell require a runtime of about 100 days when the physical error rate is 0.1\%. Reducing the physical error rate by an order of magnitude to 0.01\% leads to the reduction of the runtime by approximately a factor of 2 for all systems considered in this work. 

The number of physical and logical qubits for simulations described above are shown in Fig.~\ref{fig: Runtime_Logical_Phys}(b,c). As can be seen from Fig.~\ref{fig: Runtime_Logical_Phys}(b), the smallest simulations will require few million physical qubits if the physical error rate reaches 0.01\%, while the largest simulations of NiO and PdO need about 65 million physical qubits. For the error rate of 0.1\%, quantum error correction requires the number of physical qubits to be 4-5 times larger. The number of logical qubits [see Fig.~\ref{fig: Runtime_Logical_Phys}(c)] required for the simulation of small cells is around few thousands while large super cells would need around $10^{5}$ logical qubits. We note that the improvement in the physical error rate occasionally reduces the number of logical qubits. This is because the number of logical qubits is the sum of the computational qubits and the magic state factory qubits.  The computational qubit count is a feature of the system we study, and does not depend on the error rate of the quantum computer.  The number of logical qubits dedicated to magic state distillation could in principle change when the error rate changes -- because the fidelity with which we need to distill magic states depends on the error rate.  Yet, many of our resource estimations just require the highest fidelity factory (225-to-1) in our list [see footnote in Sec.~\ref{sec:error_correction}] for either error rates (0.1\% or 0.01\%), and hence the logical qubit count does not change.  For much smaller error rates, smaller factories would suffice and we would see such changes.

\section{Discussion and Conclusion}\label{sec:disc_and_concl}

The estimation of the ground state energy of crystalline solids with a supercell of {\it ca.}~50--70 atoms requires {\it ca.}~$10^{10}$--$10^{12}$ T gates when the size of the basis set is {\it ca.}~300--500 spatial orbitals. This is comparable to the T-gate count required for the estimation of the energy of molecular systems within the active space of several tens of orbitals. For example, simulation of a Ru complex with 65 spatial orbitals and using double factorization (DF) requires around $4.6 \times 10^{10}$ T gates~\cite{von_burg_quantum_2021} while simulation of cytochrome P450 with 58 spatial orbitals and using tensor hypercontraction (THC)~\cite{lee_even_2021} requires around $7.8 \times 10^9$ T gates~\cite{goings2022reliably}. Thus, if molecular Hamiltonians can be simulated within a reasonable time then so can a Hamiltonian describing crystalline solids.
However, the number of logical qubits is larger for solids, $10^{4}$--$10^{5}$, which is due to the fact that the number of orbitals considered in this work is larger almost by an order of magnitude as compared to molecular resource estimates.

We have considered the use of minimal Gaussian basis sets. However, for realistic solids one would need to use at least DZP or TZP basis sets, which would lead to a higher T-gate count. However, numerical studies for molecular systems indicate that methods like THC provide the best asymptotic scaling with respect to the number of orbitals for molecules~\cite{goings2022reliably}. Thus, using such approaches one might still obtain a reasonable resource estimates for crystalline solids. In order to carry out such estimations, one would have to (i) adapt such methods for periodic systems by also taking the translation symmetry into the consideration and (ii) develop classical electronic structure software for efficient generation of factorized Hamiltonians. For example, in order to generate ERIs for NiO with 64 atoms in the supercell and using GTH-DZP basis set one would need to use several TB of memory. Another approach which would allow to perform useful simulations of solids on error-corrected quantum computers with larger basis sets is by choosing the active space within the size of several hundred of orbitals or by using quantum embedding methods. 

In this work, we have investigated how the translational symmetry of the Hamiltonian can be exploited in order to reduce the quantum resources. Similarly, other symmetries such as point group symmetry can be taken into account. However, we do not expect that this will lead to an order of magnitude reduction of T gates as qubitization-based algorithms scale as the square root of the number of terms. The use of Brillouin zone symmetry can also reduce the cost of quantum algorithms in Bloch basis set but we expect that Wannier representation will provide a better resource estimates for moderate-size systems. For example, 27 $k$-points were used for NiO and even if each $k$-point provided the same coefficients it would reduce the cost by a factor of 5.

The efficiency of the QPE to estimate the ground state energy depends also on the overlap between an initial state such as Hartree-Fock state and the true ground state wave function. We have not investigated this in this work. For molecular systems, this overlap can appear to be either sufficient~\cite{von_burg_quantum_2021,goings2022reliably} or small~\cite{lee2022there} and more investigation needs to be carried out for crystalline solids. In this work, we have focused only on the single-shot cost of the total QPE circuit. 

In classical computations of ground state energy of periodic solids, using  Bloch functions is currently considered to be the most efficient approach in both KS-DFT~\cite{Kresse1996} and wave function methods such as coupled-cluster theory~\cite{PySCF4}. Wannier functions are often used as a post-processing tool for calculation of properties such as conductivity or band structure interpolation~\cite{Marzari2012}. In quantum computing, however, the Wannier functions represent an efficient choice as a basis set for the ground state energy calculations. Other areas, such as linear-scaling DFT also uses the Wannier functions~\cite{Skylaris_2002,Mostofi_2002} as the primary basis set.

In conclusion, we have considered ground state energy estimation of crystalline solids on error-corrected quantum computers using qubitization based QPE. We present two materials such as NiO and PdO which are known to be challenging systems for electronic structure methods on classical computers and are relevant for heterogeneous catalysis. The investigation of properties beyond ground state energy calculations will be addressed in the future; this work is a first step towards practical algorithms for simulation of crystalline solids on error-corrected quantum computers. We have adapted the qubitization algorithm to solid-state systems by taking into account the symmetries of the integrals in Wannier representation and generalized sparse qubitization for use with complex Hamiltonians which are needed when Bloch functions are employed. Realistic resources estimations have been carried out and presented for error-corrected quantum computers. The simulation of crystalline solids in the minimal basis set on a quantum computer with the approach presented in this paper would require an order of 10--100 millions of physical qubits and $10^{10}$--$10^{12}$ number of T gates. We expect that these numbers can be reduced further by using, for example, different qubitization techniques such as DF~\cite{von_burg_quantum_2021} or THC~\cite{lee_even_2021} adapted for solid state systems.

\section{Acknowledgment}
We thank Robert {\'I}szak for valuable discussion and comments on the manuscript. We also thank Nick Blunt for helpful discussion as well as Earl Campbell for useful discussion on algorithms and error correction.

\section{Supplemental Information}
The geometry and integral files for all systems considered in this work are available at Zenodo~\cite{SI_files}.

\bibliography{main}

\appendix

\section{Detailed costings}
\label{sec:detailed costings}

In this appendix, we give detailed costings for our Bloch and Wannier algorithms that include subdominant contributions. Since the two algorithms are very similar, we present them together and will note wherever there is a difference. All logarithms are base 2 as they refer to necessary qubit count for an integer range.
The costings of standard circuit elements follow the costings of the sparse algorithm in Appendix A of Ref.~\cite{lee_even_2021}.

\subsection{Parameters}
\begin{itemize}
\item Error parameters $\epsilon_{\rm QPE}, \epsilon_{prep}, \epsilon_{trunc}$ are discussed in sec.~\ref{sec:errors and truncation}.
\item $d$ is the number of non-zero coefficients in the Hamiltonian's LCU, after truncation and up to the symmetries restored. This means in the Wannier basis, one has to divide the number of non-zero coefficients (not taking into account translation symmetry) by the total number of cells $N$.
\item $\lambda$ is the $L_1$ norm of the LCU which is the the normalization factor of the block-encoded Hamiltonian. See \eqref{eq:lambda}.
\item $M$ is the number of spatial orbitals per unit cell.
\item $N_1,N_2$ and $N_3$ are the number of unit cells in each direction of the lattice as discribed int he main text. Therefore, the total number of spatial orbitals in the crystal is $P = MN_1N_2N_3$ and the number of spin orbitals is $2P$.
\item $\aleph$ is the size of the ``keep" register in coherent alias sampling. Coherent alias sampling effectively sets the amplitudes of the state based on the $\aleph$ register. Therefore it affects the accuracy of the prepared state; following (A12) \cite{lee_even_2021},
\begin{equation}\label{eq:aleph}
\aleph = \left\lceil \log \frac{\lambda}{2\epsilon_{prep}}\right\rceil.
\end{equation}
\item $m$ is the output size (qubit number) of the QROAM, the data loading in PREPARE.
For Bloch functions, it is
\begin{equation}
m =\aleph + 2(4\lceil \log (P)\rceil + 6)
\end{equation}
and for Wannier functions,
\begin{equation}
m = \aleph + 2(4\lceil\log M\rceil + 3 \sum_{i=1,2,3} \lceil\log N_i\rceil + 4),
\end{equation}
where the compound index register must be split into its constituents. The $+6$ (Bloch) or $+4$ (Wannier) is made up of one qubit to distinguish one- and two-body terms, one qubit for the sign of the coefficient, three qubits (Bloch) or one qubit (Wannier) identifying term types for Majorana type restoration and one qubit for spin restoration. (Note that contrary to the simplified depiction in Figs.~\ref{fig:prepare bloch} and \ref{fig:prepare wannier}, the $\ket{\sigma}$ qubit in \eqref{eq:spin restore table} is added as an ancilla rather than loaded.) The factor of 2 stems from needing ``ind" and ``alt" values for coherent alias sampling.
\item $\kappa_1$ and $\kappa_2$ are powers of 2. They determine the space-time tradeoff in the QROAM and QROAM uncompute. We choose them to minimize Toffoli cost.
\item $\mathcal{I} = \left\lceil \frac{\pi\lambda}{2\epsilon_{\rm QPE}}\right\rceil$ the number of repetitions of the walk operator for quantum phase estimation
\item $2^\eta$ is the maximal power of 2 that's a factor of $d$
\item $b_r$ are bits of precision for the equal state preparation, we take it 7 as suggested in Ref.~\cite{lee_even_2021}.

\end{itemize}

\subsection{Toffoli count}
The total Toffoli count is the product of $\mathcal{I}$ (the number of iterations in phase estimation) and the the number of Toffolis needed to construct the walk operator. The walk operator consists of the following circuit elements:
\begin{itemize}
\item PREPARE and UNPREPARE. The PREPARE operator is sketched in Figs.~\ref{fig:prepare bloch} and \ref{fig:prepare wannier}, while UNPREPARE uncomputes the state. We state the total cost of both, leading to a factor of two in most items.
\begin{itemize}
    \item Equal state superposition over $d$ basis states via amplitude amplification:
    \begin{equation}
    2(3\lceil\log d\rceil - 3\eta + 2 b_r -9)
    \end{equation}
    \item Data lookup via QROAM. This is the asymptotically dominant contribution to the walk operator. Uncomputing can be significantly simplified using a measurement based uncomputation scheme, such that the total cost is:
    \begin{equation}
    \lceil d/\kappa_1\rceil + m(\kappa_1-1) +\lceil d/\kappa_2\rceil +\kappa_2
    \end{equation}
    \item Coherent alias sampling. It can be uncomputed without Toffolis, giving:
    \begin{equation}
    \aleph + (m-\aleph-2)/2.
    \end{equation}
    The $\aleph$ is for an inequality test, and the register sizes to be swapped are $(m-\aleph)/2$. Swapping the sign qubits can be done without Toffolis, leading to the $-2$.
    \item Majorana type symmetry restoration circuit \eqref{eq:maj type restore circuit} for Bloch basis:
    \begin{equation}
    2(5 + 25)
    \end{equation}
    This cost includes a unary iteration, controlled on the 1body/2body qubit and iterating over six values of the $\ket{s_i}$ qubits (5 Toffolis).
    For the Wannier basis, the circuit can be simplified and the cost is only
    \begin{equation}
    2(1 + 7).
    \end{equation}
    A controlled Hadamard can be implemented with a single Toffoli using a catalytic T state, see Fig.~17 in Ref.~\cite{lee_even_2021}.
    \item Spin symmetry restoration: 0
    \item Translational symmetry restoration (Wannier basis only). The cost of computing and uncomputing the additions in \eqref{eq:wannier in-place additions} is:
    \begin{equation}
    2\left( \sum_{i=1,2,3}3(\log N_i -1)\right). \label{eq:cost additions}
    \end{equation}
    For this formula we have assumed that $N_1, N_2, N_3$ are all powers of 2. Then the in-place additions (or in the uncomputation, subtractions) modulo $N_{i}$ can be performed with $\log N_{i} -1$ Toffolis each \cite{Gidney2018Halving}. If $N_{i}$ is not a power of 2, the cost will be higher to perform the correct modular arithmetic. Yet this symmetry restoration circuit is a subleading contribution and for simplicity we use \eqref{eq:cost additions} even if the cell numbers are not powers of 2.
\end{itemize}
\item SELECT. We have two ranged operations that are uncontrolled and two that are controlled on the qubit flagging one- or two-body terms and thereby the number of Majoranas in the unitary. Each of these unary iterations is over $4P$ values. These are spin, Majorana type, and the indices specifying each Majorana [See Figs.~\ref{fig:prepare bloch}, \ref{fig:prepare wannier}]. The total cost is
\begin{equation}
2(4P-2) + 2(4P-1).
\end{equation}
\item Reflection. The walk operator is built from the block-encoded Hamiltonian along with a reflection. This is implemented by a multicontrolled $Z$ controlled on a  number of qubits. These are $\lceil\log d\rceil$ for the QROAM index, one further qubit from preparing the equal superposition state, $\aleph$ for the equal superposition state in coherent alias sampling, one qubit from spin, four qubits from Majorana type, and (only for Wannier) $\lceil\log N_1 \rceil + \lceil\log N_2 \rceil + \lceil\log N_3 \rceil$. The gate controlled on $c$ qubits can be implemented with $c-1$ Toffolis, giving
\begin{equation}
\lceil \log d\rceil + \aleph + 5 + (\text{Wannier only:}\ \sum_{i=1,2,3} \lceil\log N_i \rceil)
\end{equation}
Toffolis.
\item Two more Toffolis for each step (to make the reflection controlled, and for the unary iteration in the phase estimation)
\end{itemize}

Note we do not include the cost of initial state preparation for the Heisenberg-limited phase estimation or of the inverse QFT. These are small additive costs that are not multiplied by $\mathcal{I}\propto \lambda/\epsilon_{\rm QPE}$ in contrast to all the above contributions.

\subsection{Qubit count}
\begin{itemize}
\item For phase estimation and unary iteration circuit:
$\lceil\log(\mathcal{I}+1) \rceil + \lceil\log(\mathcal{I}+1)\rceil -1$
\item System qubits (on which the Majoranas act):
$2P$
\item QROAM input state: $\lceil\log d\rceil$
\item $\aleph + 2 + b_r$ for equal superposition state and coherent alias sampling
\item QROAM qubits including output and ancillas: $m\kappa_1 +\lceil\log(d/\kappa_1)\rceil$
\item Further ancillary qubits are not needed as QROAM ancilla qubits can be reused.
\end{itemize}

\end{document}